# Unleashing the power of text for credit default prediction: Comparing human-written and generative AI-refined texts

Zongxiao Wu[a], Yizhe Dong[a, *], Yaoyiran Li[b], Baofeng Shi[c]

[a] Business School, University of Edinburgh, Edinburgh, EH8 9JS, United Kingdom
[b] Language Technology Lab, University of Cambridge, Cambridge, CB3 9DA, United Kingdom
[c] College of Economics and Management, Northwest A&F University, Yangling, 712100, China

**Abstract**

This study explores the integration of a representative large language model, ChatGPT, into lending decision-making with a focus on credit default prediction. Specifically, we use ChatGPT to analyse and interpret loan assessments written by loan officers and generate refined versions of these texts. Our comparative analysis reveals significant differences between generative artificial intelligence (AI)-refined and human-written texts in terms of text length, semantic similarity, and linguistic representations. Using deep learning techniques, we show that incorporating unstructured text data, particularly ChatGPT-refined texts, alongside conventional structured data significantly enhances credit default predictions. Furthermore, we demonstrate how the contents of both human-written and ChatGPT-refined assessments contribute to the models' prediction and show that the effect of essential words is highly context-dependent. Moreover, we find that ChatGPT's analysis of borrower delinquency contributes the most to improving predictive accuracy. We also evaluate the business impact of the models based on human-written and ChatGPT-refined texts, and find that, in most cases, the latter yields higher profitability than the former. This study provides valuable insights into the transformative potential of generative AI in financial services.

* Corresponding Author.
Email addresses:
Zongxiao.Wu@ed.ac.uk, Yizhe.Dong@ed.ac.uk, yl711@cam.ac.uk, shibaofeng@nwsuaf.edu.cn

> *"Although humans are still better than GPT at a lot of things, there are many jobs where these capabilities are not used much. […] and soon these data sets will also be used to train the AIs that will empower people to do this work more efficiently"*.
>
> -- *The Age of AI has Begun* by Bill Gates[1]

# 1. Introduction

Since its launch in November 2022, OpenAI's ChatGPT, powered by a large language model (LLM), has attracted over 180 million users. This success has triggered a global wave of innovation in the field of artificial intelligence (AI). Recent advancements in generative AI are significantly influencing industry sectors, altering both business operations and information acquisition methods in the digital era. This technology can generate novel content, including text, images, audio, simulations, and synthetic data, by analysing patterns and structures within existing data. We have seen an explosion in applications of generative AI for software, marketing, gaming, e-commerce, healthcare, and education companies. However, its potential within the financial sector remains largely untapped. Financial institutions and investors still lack sufficient knowledge and understanding to make informed decisions regarding the incorporation of generative AI into investment and lending strategies. Therefore, this paper aims to bridge these practical challenges by studying the adoption of generative AI tools and natural language processing (NLP) techniques in the lending decision-making process and thereby show the importance of leveraging alternative data to supplement traditional credit data.

Over the last decade, machine learning (ML) has played an instrumental role in propelling the lending industry forward, enabling notable advancements such as improved default prediction, better underwriting, and greater financial inclusion. Many studies have adopted ML techniques such as neural networks, decision tree (DT), random forest (RF), XGBoost, and support vector machine (SVM) for default prediction and have shown that they are superior to traditional statistical methods. In the era of big data, several recent studies have sought to leverage text data, a major source of alternative data, for credit default prediction (e.g., Jiang et al., 2018; Stevenson et al., 2021; Kriebel & Stitz, 2022). They utilise NLP tools such as word embedding models, convolutional neural networks (CNNs), and transformer models to extract insights from text data and demonstrate that integrating this unstructured data into the modelling process can significantly improve predictive performance. However, most of these studies have centred around peer-to-peer lending. There is limited evidence pertaining to traditional bank loans and exploiting the recent advances in LLMs in financial and business contexts.

In this study, we attempt to explore the effectiveness of a representative LLM, ChatGPT, for analysing, distilling, and refining information from loan-related text within the context of credit default prediction. We believe that using generative AI tools like ChatGPT to refine human-written loan assessments can offer benefits to the risk assessment process. First, human-written loan assessments often vary significantly in terms of structure, terminology, length, and complexity due to individual writing styles and subjective interpretations. This inconsistency introduces noise and makes it challenging for NLP models to then effectively extract relevant information. ChatGPT can process these diverse texts and produce summaries or refinements that follow a consistent format, logic, and standardised language (Wu et al., 2024). Moreover, recent studies have shown that LLMs possess strong capabilities in text understanding and generation (Xu et al., 2024; Achiam et al., 2024). These models excel at discerning context from natural language, allowing them to extract essential information from human-written texts without extensive task-specific training. By utilising ChatGPT to refine human-written loan assessments, we anticipate that it can effectively identify and emphasise critical credit risk information while filtering out irrelevant details. Lastly, several studies have demonstrated that large-scale LLMs with over 100 billion parameters exhibit human-like reasoning ability (Dasgupta et al., 2023). This capability enables LLMs to uncover hidden information and infer additional relevant cues for downstream tasks that require more complex and nuanced thinking, such as credit default prediction. Huang & Chang (2023) suggest that this could improve the performance of subsequent NLP models and enhance their robustness in handling out-of-distribution data, a common challenge in predictive modelling. Therefore, this paper aims to empirically investigate the potential of ChatGPT for enhancing the accuracy and effectiveness of credit default prediction. Specifically, we endeavour to examine whether integrating insights from textual data into credit scoring models can significantly improve loan default prediction, and if so, which type of loan assessments, human-written or ChatGPT-refined, exhibits greater predictive effectiveness?

---

[1] *The Age of AI has begun* by Bill Gates. https://www.gatesnotes.com/The-Age-of-AI-Has-Begun

Our empirical analyses are based on a unique loan dataset containing 2,460 micro and small business (mSE) loans, featuring both structured and text data. We utilise ChatGPT to analyse the original human-written loan assessments and then derive AI-refined analytical assessments. Through comparisons between these two types of texts, we find that human-written and ChatGPT-refined texts are quite different in terms of text length, semantic similarity, and linguistic representations. In the modelling part, we develop text-enhanced credit scoring models that leverage human-written and ChatGPT-refined texts, respectively, based on state-of-the-art NLP models. While both texts benefit credit scoring models (compared to models using only structured data), ChatGPT-refined texts can lead to more accurate prediction results in most cases. Moreover, we investigate which particular signal words and phrases within each text type strongly contribute to predictive capacity. Using local interpretable model-agnostic explanations (LIME), we find that the effect of essential words on prediction results depends on the context in which they appear. Furthermore, we explore whether all sections of ChatGPT-refined text carry equal potential for predicting defaults. The results suggest that ChatGPT demonstrate robust capabilities in identifying a borrower's delinquency factors by assessing the information contained in the original loan assessments. In addition, we use several alternative generative AI tools for comparative analysis. These results further reinforce our findings and highlight the efficacy of GPT-based models in analysing textual information. To evaluate the business impact, we also estimate the monetary gains associated with models based on human-written versus ChatGPT-refined texts. The results indicate that, in most cases, the model relying on ChatGPT-refined texts yields higher profits compared to that using human-written texts.

Our study makes several contributions to the field. First, our work benefits the fast-growing research on generative AI applications in finance and business alike. While previous works have demonstrated that ChatGPT can facilitate stock return prediction (Lopez-Lira & Tang, 2023; Kim et al., 2023), our paper attempts to examine the potential of generative AI for enhancing lending decision-making. Our results suggest that AI-refined texts add considerable value to loan default prediction as they can amplify useful information in human-written textual assessments. Our paper also compares the efficacy of multiple generative AI models, which provides actionable insights for institutions and organisations that intend to embrace generative AI tools to improve their financial decision-making. Second, our study complements and contributes to the existing literature on the value of alternative data, especially generative AI-refined text data, for predicting important outcomes in financial markets, including voice (Yang et al., 2023), visual content (Bazley et al., 2021), Google search data (Da et al., 2015), and satellite data (Mukherjee et al., 2021). Our research demonstrates that including such texts can significantly improve loan default prediction for mSEs whose structured data (i.e., financial information) are often unavailable to lenders. Third, our research comprehensively investigates the linguistic differences between loan assessment texts produced by humans and LLMs. Our results indicate that these two types of texts are significantly different concerning text length, semantic similarity, and linguistic representations.

The remainder of the paper is organised as follows. In Section 2, we discuss prior literature on NLP-based default prediction and recent LLM applications in the finance and business sectors. Section 3 provides a detailed description of our dataset and the ChatGPT prompt proposed for generating new content. We also investigate the linguistic differences between human-written and ChatGPT-refined texts. Section 4 outlines the modelling process and methods employed in this study. In Section 5, we present our prediction results and analyse how text influences the model performance. We also provide further analysis of the default prediction experiments carried out using texts refined by alternative generative AI tools. Section 6 evaluates the business impact of the models based on human-written and ChatGPT-refined texts. Finally, we summarise our work's key findings and implications in Section 7.

**2. Literature review**

Here, we organise the prior literature into two main categories and point out the research gaps that motivate our work. Section 2.1 briefly summarises the use of NLP techniques in credit default prediction; Section 2.2 reviews recent studies exploring LLMs' potential in finance and business sectors.

*2.1 NLP-based default prediction*

Credit scoring refers to an automatic credit assessment procedure that helps lenders identify borrowers who will fail to fulfil their financial obligations within a group of loan applicants. Given the surge in the consumer lending market, novel credit scoring models continue to attract considerable interest in academia and industry (Song et al., 2023). Although traditional credit scoring models have primarily relied on structured data, strong evidence shows that integrating text data can significantly enhance prediction results through the leveraging of modern NLP techniques (Wang et al., 2020; Kriebel & Stitz, 2022).

Early credit scoring studies explored the value of text using traditional text mining approaches. For example, Herzenstein et al. (2011) highlighted the significant influence of identity claims such as 'trustworthy', 'successful', and 'moral' extracted from loan narratives on lending decisions. Iyer et al. (2016) derived several characteristics from borrowers' loan descriptions, including average word length and percentage of misspelt words, among others, to predict loan default. Although these approaches can, to some extent, exploit the relationship between texts and human cognitive processes, their application requires expert knowledge and significant manual efforts (Kriebel & Stitz, 2022). As digital technologies flourished, scholars started to employ statistical NLP models to extract textual features. Pioneering these approaches, Jiang et al. (2018) used latent Dirichlet allocation (LDA), a widely used topic model, to derive features from borrowers' written descriptions. Fitzpatrick & Mues (2021) applied a biterm topic model to generate text representations for loan descriptions, thereby addressing the sparsity problem of short texts. Other statistical methods, including latent semantic analysis and term frequency-inverse document frequency (TF-IDF), have also been widely used in credit scoring studies (Netzer et al., 2019; Xia et al., 2020). However, while these statistical techniques are easy to scale, they may fail to comprehend complex linguistic phenomena and suffer from the curse of dimensionality.

Given the breakthroughs in deep learning in various fields, scholars have increasingly explored the potential of neural NLP models for extracting textual information. Earlier works adopted static word representations such as global vectors for word representation (GloVe) and fastText to transform words into real-valued vector representations which could be used for downstream tasks with subsequent classifiers. In the context of default prediction, Mai et al. (2019) pioneered the use of CNNs, while Matin et al. (2019) simultaneously leveraged recurrent neural networks to predict corporate bankruptcies, gleaning insights from audit reports and management statements; Wang et al. (2020) and Kriebel & Stitz (2022), however, adopted more traditional classifiers such as logistic regression (LR) and RF to predict borrowers' defaults by leveraging GloVe embeddings. More recently, pre-trained language models have revolutionised the field of NLP, offering notable performance enhancements: neural networks are initially pre-trained on massive amounts of human language data and then adapted to various downstream tasks through task-specific fine-tuning (a method of transfer learning). Transformer models, such as bidirectional encoder representations from transformers (BERT) and the robustly optimised BERT pre-training approach (RoBERTa), are now state-of-the-art across many benchmark NLP tasks. While these models are well-known in NLP and AI communities, it was not until 2021 that they were introduced for credit default prediction. Stevenson et al. (2021) shed the first light on utilising BERT to derive text representations from loan officers' assessments for bankruptcy prediction. Subsequently, Fitzpatrick & Mues (2021) applied BERT, embeddings from language models (ELMo), and the universal sentence encoder (USE) to capture credit-relevant information contained in borrowers' loan descriptions; with BERT and RoBERTa, Kriebel & Stitz (2022) showed that even succinct text descriptions of borrowers could considerably enhance default prediction. More recently, Xia et al. (2023) and Sanz-Guerrero & Arroyo (2024) innovated by incorporating default probabilities derived from fine-tuned BERT and RoBERTa as textual features in credit scoring models, showcasing that narrative data contain valuable credit information.

*2.2 LLMs' applications in finance*

The emergence of LLMs can be traced back to the development of transformer architecture that exploited the attention mechanism to model long-range dependencies in a sequence (Vaswani et al., 2017). Transformer-based LLMs with a huge number of parameters (e.g., billions) are trained on massive amounts of text data from diverse sources, and they can be adapted to a broad spectrum of downstream tasks. Evidence suggests that LLMs achieve impressive results over a range of NLP tasks, including question answering, machine translation, and recommender systems (Wu et al., 2023; Li et al., 2024a; Li et al., 2024b). In November 2022, the release of OpenAI's ChatGPT catapulted LLMs to the centre of the NLP community and attracted a significant amount of discussion worldwide. Building on existing GPT infrastructure, ChatGPT fuses unsupervised learning with reinforcement learning derived from human feedback, generating responses to inquiries that closely mimic human discourse. Given the unprecedented capabilities of LLMs, such as ChatGPT, some studies have started to explore their potential use in various domains, but there are few relevant studies in the financial field (Lopez-Lira & Tang, 2023).

Most of the existing literature provides potential application scenarios for the use of LLMs within the financial domain, taking ChatGPT as a core example. Wenzlaff & Spaeth (2022) first explored how ChatGPT defines concepts such as crowdfunding, alternative finance, and community finance, and how these concepts correspond to human answers in academic scholarship. Dowling & Lucey (2023) suggested that ChatGPT could generate plausible-seeming financial research studies and that the results could be significantly improved by integrating private data and researcher expertise. Cao & Zhai (2023) presented several examples of how to guide ChatGPT to generate outcomes for financial and accounting research questions, including sentiment analysis, ESG (environmental, social, and governance) analysis, and corporate culture analysis. Alshurafat (2023) reviewed the potential use of ChatGPT in financial and accounting applications and discussed ethical considerations.

Other literature has delved into using LLMs in financial market applications by leveraging quantitative methodologies, especially in equity investment. For example, Lopez-Lira & Tang (2023) examined the capabilities of various LLMs to predict stock returns via sentiment analysis of news headlines. Their findings suggest that ChatGPT outperforms traditional models, and that integrating such advanced LLMs into investment decision-making could lead to more accurate prediction results and enhance the performance of trading strategies. Yue et al. (2023) applied an explainable AI model to ChatGPT to interpret the stock performance generated by non-linear ML models. In another study, Ko & Lee (2023) demonstrated the efficacy of ChatGPT in portfolio management, particularly in asset allocation and diversification. Their analysis shows that the portfolios selected by ChatGPT are notably superior, in terms of a diversity index, to randomly selected assets. Additionally, Kim et al. (2023) used ChatGPT to distil textual information disclosed in companies' annual reports and conference calls and found that the ChatGPT-generated content is more effective in explaining stock market reactions.

In the context of credit scoring, several recent studies have investigated the potential of LLMs for accurately predicting binary outcomes. This approach involves incorporating structured data (tabular information) with textual inputs into the prompt. Deldjoo (2023) demonstrated that, with carefully designed prompts enriched with domain-specific knowledge, ChatGPT could outperform traditional ML models while requiring significantly less training data. Similarly, Babaei & Giudici (2024) applied GPT-3.5 using few-shot learning for lending decisions and showed that GPT could achieve results comparable to traditional credit application models. In addition, Yu et al. (2023) employed ChatGPT to extract psychological features, specifically the big five personality traits, from borrowers' loan descriptions. This extracted information was then integrated with structured data in subsequent classifiers. They found that their models offer a robust solution for credit scoring, balancing predictive accuracy with data privacy. Furthermore, Feng et al. (2024) curated an instruction tuning dataset tailored to the application of LLMs in credit assessment and demonstrated that LLMs surpass conventional models in credit scoring. They also emphasised the necessity for careful consideration of fairness and ethical practices when deploying LLMs in the financial industry. Lastly, Li et al. (2024) conducted a fairness evaluation of ChatGPT in credit risk management by evaluating the model's responses to both biased and unbiased prompts. Their findings suggest that, while ChatGPT demonstrated improved fairness compared to smaller models like LR and multilayer perceptron (MLP), it still exhibited some inherent biases influenced by the prompt engineering.

In summary, our review of existing studies reveals the importance and effectiveness of using NLP techniques to extract text features for default prediction, and there is indeed growing interest in applying LLMs to address problems in the finance and business sectors. While some studies have explored the integration of LLMs into the making of informed financial decisions, large gaps still remain in effectively utilising these models to augment lending strategies and improve credit risk management. This paper aims to bridge these research gaps, offering insights that could potentially benefit a wide range of financial and business applications.

## 3. Data description and collection

The mSE loan dataset used in this study was provided by a Chinese bank operating on a national basis. In 2008, the bank launched a specialised product to support the short-term finance needs of small businesses. The dataset originally includes 3,000 commercial and industrial loans, primarily granted to mSEs for immediate operating needs, such as managing cash flow, financing inventory purchases, or covering unforeseen expenses. These loans are attractive to small businesses because they can be obtained more quickly than long-term financing options and offer flexibility for managing short-term financial needs without the burden of extended debt commitments. After removing records lacking textual information and those with significant missing variables, the final sample consists of 2,460 loans with durations ranging from 1 to 8 months, originating between December 2008 and July 2009. All the loans are closed – 2,400 loans were fully paid off at maturity or early, and 60 borrowers defaulted.[2] Default is defined as 90 days past payment due. Each borrower has 18 nominal and 13 numeric attributes, one text extracted from the loan assessments generated by a loan officer, and a final class label (i.e., defaulter or non-defaulter). Besides the original text, we also derive a new text from ChatGPT-refined analytical assessments.

---

[2] We note that the dataset is characterised by a severely imbalanced ratio of 1:40 (defaulter: non-defaulter). However, this was quite common among Chinese lenders during this time and remains so today. For example, according to the China Banking and Insurance Regulatory Commission, the average non-performing loan ratio for the Chinese banking sector was 1.58% in 2009.

*3.1 Structured features*

The original dataset comprises 31 standard credit features, including information on borrowers' demographics, business details, and loan particulars. In cases where continuous features are missing, they are filled with the means of the features, while missing values for categorical features are replaced with new categories. Appendix Table A.1 in the supplementary materials presents the definitions and summary statistics of all standard variables in our dataset. To ensure consistency and comparability of input features across models, we apply the weight of evidence (WoE) method to the features beforehand (Jiang et al., 2019). The WoE value in each category/bin for a given feature is calculated as the logarithm of the proportion of non-defaulters to the proportion of defaulters in that particular category/bin. A large negative value corresponds to a higher default risk and vice versa (Yap et al., 2011). We also employ the information value (IV) and variance inflation factor (VIF) measures to remove redundancy, reduce the multicollinearity of the input features and select an appropriate subset of features for inclusion in our credit scoring models. We calculate the IV as the weighted sum of the WoE values for each category/bin of the feature, which serves to represent the feature's predictive power. In this study, we retain those WoE-encoded features with an IV greater than 0.01 and less than 0.50. Meanwhile, we use the VIF to detect multicollinearity issues among the WoE-encoded features and only retain those features with a VIF less than or equal to 10. Through these preprocessing steps, we identify 18 structured variables for our subsequent modelling.[3]

*3.2 Human-written text*

In many cases, mSEs lack a credit history and struggle to provide necessary and reliable financial information. To obtain a comprehensive understanding of an mSE's operating and financial situation, loan officers from the lender usually conduct site visits. These visits help verify the accuracy of the information provided in the loan applications and allow the loan officers to gather first-hand information about the borrower's business, day-to-day operations, and willingness to repay, among other things (Stevenson et al., 2021). Based on the information gathered, these officers then prepare loan assessment reports, helping the loan underwriters assess the potential credit risk of the borrowers and make informed decisions on loan approvals. In our dataset, the loan assessment typically evaluates the borrower from three perspectives: an overall impression of the borrower, the borrower's repayment intentions, and the borrower's assets and liabilities. The loan underwriters would then thoroughly review the loan assessments alongside the borrowers' applications, financial information, and any available credit histories to decide whether to approve or reject loan applications. To ensure the text is clear for our analysis, we remove redundant white spaces in the text and add a full stop at the end of each complete sentence, steps which can enhance a text's readability. Below, we present a typical extract from the defaulters' textual loan assessments, which has been translated into English from the original Chinese. This extract has been fully anonymised, with all quantitative values masked. For comparison, a typical extract from the non-defaulters' textual loan assessments is available in Appendix B of the supplementary materials.

---

[3] As a robustness check for the feature selection, we further implemented the one-way analysis of variance (ANOVA) and backward stepwise feature selection methods to select structured features (Hajek & Michalak, 2013). The unreported results based on different subsets align with our initial findings, emphasising the stability and efficacy of our feature selection approach.

Human-written text for a defaulter: *"The borrower has good peer relationships, actively cooperates with the credit officer's investigation, and provides valid documentation. The borrower has marketing work experience and a certain understanding of the industry's business environment. The borrower knows the consequences of loan default, and no obvious factors might affect his willingness to repay. However, through an individual credit check, the credit officer found that the borrower had a history of several late credit card payments caused by his being busy with work and not making timely payments, thus reducing his credit rating to A. The borrower's cash and bank balance were approximately ¥ ???, and the borrower's current receivables were approximately ¥ ???, which is consistent with the borrower's statements. The inventory in the borrower's shop is valued at the purchase price of approximately ¥ ???."*

As illustrated in Figure 1, the text lengths for the Goods (non-defaulters) group appear to follow a symmetrical Gaussian distribution, while those of the Bads (defaulters) group roughly exhibit a right-skewed distribution. There are notable differences between the two groups. The Bads group exhibits a lower mean word count of 164 and a larger standard deviation (SD) of 93. In contrast, the Goods group displays a higher mean word count of 210 and a smaller dispersion (SD 75). These findings suggest that loan officers tend to generate more extended and detailed texts for loans deemed safe, while providing more concise assessments for risky loans. One possible explanation for this pattern is that non-defaulters tend to provide more positive attributes, while both non-defaulters and defaulters downplay or conceal potential risk factors. When borrowers emphasise their positive attributes, this may lead loan officers to focus more on the strengths of the loan application, resulting in more extended assessments. Conversely, when borrowers provide little information, loan officers may face difficulties uncovering hidden risk factors, which can contribute to more concise loan assessments. By recognising the tendencies in the text lengths and levels of detail in loan assessments, we can gain insights into the assessment practices of loan officers and obtain potential implications for predicting default risk.

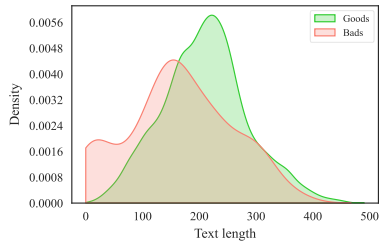 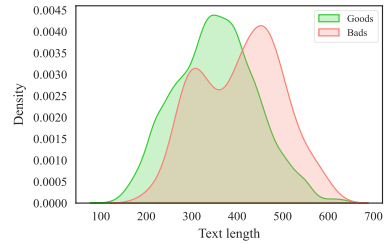

**Figure 1** Text length of human-written texts    **Figure 2** Text length of ChatGPT-refined texts

*3.3 ChatGPT-refined text*

In this study, we employ ChatGPT[4] to analyse human-written loan assessments and generate AI-refined analytical assessments. To enhance ChatGPT's understanding of the task and instruct it to generate the desired responses, we adopt a well-designed prompt template which is set up to be clear and specific, and containing sufficient background. Our prompt template consists of two main components: (1) the initial loan assessment about the borrower provided by the loan officer, and (2) an outline of the task that we want ChatGPT to perform. Specifically, we instruct ChatGPT to identify factors that support a borrower's likelihood to repay and factors that may lead to the borrower's default. These are crucial factors in the lender's decision-making process. Note that we do not impose any constraints on ChatGPT's output length. Since ChatGPT was only available through its website in February 2023, we obtained the responses by entering the prompts into the input box. The chat history can influence the answers produced by ChatGPT to a large extent, so we refreshed the thread for each question. Below, we provide the detailed prompt template employed in this study for each original human-generated loan assessment:

Prompt template: *"Hi ChatGPT, there is a bank loan borrower whose details are below." + [Insert the original texts] + "The bank plans to lend to this borrower. Based on the above information, please carefully summarise and analyse the factors that support the borrower's ability to repay the loan on time and the factors that could lead to the borrower's default. The expected answer template consists of two parts: 1. Factors supporting the borrower's repayment: [Insert answer here]; 2. Factors that could potentially lead to the borrower's default: [Insert answer here]."*

We feed the templated text into ChatGPT and retain ChatGPT's analysis of the factors that favour the borrower's repayment and those that have the potential to lead to the borrower's default. The following is the output of ChatGPT corresponding to the previous example.

ChatGPT-refined text for a defaulter: *"1. Factors supporting the borrower's repayment: * The borrower has good peer relationships, actively cooperates with the credit officer's investigation, and provides valid documents, which indicates that the borrower has a good cooperative attitude and integrity. This is conducive to their establishing a cooperative relationship with the bank. * The*

---

[4] The version of ChatGPT used in this paper is version 2.13, based on the standard ChatGPT model, as updated on 13th February 2023.

*borrower has a stable social status and some resources, which may have a positive influence on the borrower's future repayment. \* The borrower attaches importance to the risk of default and has no obvious factors that might affect their willingness to repay, which indicates that the borrower has the willingness and ability to repay, and this may help the bank to conduct risk assessments and controls. 2. Factors that could lead to the borrower's default: \* The borrower's personal credit check shows that there were several overdue payments on credit cards, and this led to a downgraded credit rating of A. Although the borrower indicates that they are usually busy with work and therefore did not make timely payments, this may have a negative impact on the borrower's credit risk assessment. Therefore, the bank needs to carefully consider the setting of the loan amount, interest rate, etc. \* The borrower has a large amount of receivables, which may affect the borrower's future cash flow and hence ability to repay. Therefore, the bank needs to conduct a detailed review of the borrower's receivables to fully examine their true business operations and consider the impact of this factor in risk control."*

Figure 2 shows the text length distributions for the Goods and Bads groups' ChatGPT-refined texts. The text lengths of the Goods group appear to follow a Gaussian distribution, with a mean value of 354 (SD 89), while those of the Bads group exhibit a bimodal distribution, with a mean value of 406 (SD 90). These results contrast with our earlier findings obtained from the human assessments, suggesting that ChatGPT tends to produce more detailed and extended assessments for borrowers who are considered riskier. Moreover, upon examining the contents of the human-written and ChatGPT-refined texts, we observe that ChatGPT not only provides a good summary of the original information but also effectively eliminates faulty wording and resolves illogical ambiguities presented in the human texts. Texts refined by ChatGPT also offer additional extended analyses and implications regarding the creditworthiness of borrowers. These findings unveil ChatGPT's robust capability to reduce noise, enhance logical structure, extract essential information, and enrich content simultaneously.

We further investigate the differences between loan assessments produced by humans and ChatGPT in terms of text length, semantic similarity, and linguistic representations. Due to space constraints, detailed results and analyses are available in Appendix C of the supplementary materials. Our findings reveal significant distinctions between the two types of assessments from these perspectives.

## 4. Experimental design

*4.1 Model specifications*

The main objective of our study is to compare the predictive capabilities of human-written and AI-refined texts by leveraging NLP approaches for credit risk analysis. Specifically, we deploy four representative NLP techniques, namely LDA, fastText, Ada-002, and BERT, to derive numerical representations from texts that can then be utilised for credit scoring models. To facilitate comparisons among techniques, we use MLP[5] as the benchmark classifier for predicting default (Stevenson et al., 2021). Our prediction models can be built based on three types of features: structured (comprising only structured features), text (comprising only text features), and combined (a combination of both structured and text features). Thus, we end up with 17 different models, as presented in Table 1.

**Table 1** Input features for various default prediction models

|  | MLP | | | |
|---|---|---|---|---|
| Structured | 7 continuous 11 categorical | | | |
|  | LDA+MLP | fastText+MLP | Ada-002+MLP | BERT+MLP |
| Text-only (Human) | 30 topics | 300-dim vector | 1536-dim vector | 768-dim vector |
| Combined (Human) | 7 continuous 11 categorical 30 topics | 7 continuous 11 categorical 300-dim vector | 7 continuous 11 categorical 1536-dim vector | 7 continuous 11 categorical 768-dim vector |
| Text-only (ChatGPT) | 30 topics | 300-dim vector | 1536-dim vector | 768-dim vector |
| Combined (ChatGPT) | 7 continuous 11 categorical 30 topics | 7 continuous 11 categorical 300-dim vector | 7 continuous 11 categorical 1536-dim vector | 7 continuous 11 categorical 768-dim vector |

LDA is an unsupervised theme model used to find natural groups of topics, which models each document by identifying the underlying themes, represented as a probability distribution over a certain set of words (Blei et al., 2003). LDA has a three-layer Bayesian structure consisting of documents, topics, and words, and its posterior distribution depends on the parameters in the previous layer. The most crucial tuning parameter for LDA that needs to be predefined is the number of topics, denoted by $n$. In this study, we train the LDA model on the training set and perform hyperparameter searches within the range of [5, 10, 15, 20, 25, 30] to determine the best $n$ for the LDA model based on MLP's prediction results on validation samples. These searches imply that LDA models with 30 topics should be utilised in our subsequent modelling.

We also derive pre-trained fastText word embeddings from two types of texts. fastText, developed by Meta AI, is trained with the efficient hierarchical Softmax and skip-gram, which primarily focuses on achieving scalable solutions for text classification and representation learning while processing large datasets quickly and accurately. Compared to other earlier works such as word2vec, fastText features a deeper understanding of the internal structure of words by modelling each word as a bag of character $n$-grams. Due to its ability to capture a word's morphology and sub-word information, fastText can also provide vector representations for out-of-vocabulary words. Following Bojanowski et al. (2017), we map text inputs to 300-dimensional vectors using pre-trained fastText embeddings, which are then utilised in our modelling process.

---

[5] MLP is considered more adept at handling text representations due to its ability to perform non-linear combinations of word/sentence embeddings (Katsafados et al., 2024).

Moreover, we include OpenAI's latest embedding model, Ada-002, to convert our texts into numerical representations. Ada-002, known as text-embedding-ada-002, is a transformer-based text representation model proposed by OpenAI in December 2022 that can produce sequence embeddings for text inputs. It outperforms all OpenAI's previous embedding models on text search, code search, and sentence similarity tasks and achieves a comparable performance on text classification. As in the use of fastText, we use pre-trained Ada-002 embeddings to generate a 1536-dimensional vector for each text input that can be utilised in our subsequent credit scoring models.

Finally, we deploy BERT to extract contextualised representations for our texts. BERT is a transformer model proposed by Google AI that is trained with masked language modelling and next sentence prediction objectives (Devlin et al., 2018). BERT contains a series of so-called transformer blocks, each consisting of a bidirectional self-attention mechanism followed by feed-forward layers. Instead of reading the text input sequentially, the bidirectional paradigm allows an entire sequence of words to 'attend' to each other, thus enabling BERT to learn the word meanings in context. BERT exhibits remarkable adaptability across benchmark NLP tasks, thanks to being pre-trained on a large corpus of texts, as well as having a strong modelling capability brought about by its hundreds of millions of trainable parameters. However, generally, BERT can only process texts with a maximum sequence length of 512 tokens. The maximum sequence length for our human-written texts is 453 tokens, whereas, for the ChatGPT-refined texts, this extends to 644 tokens. Given this, 4.39% of the ChatGPT-generated texts (i.e., texts longer than 512 tokens) will be truncated during BERT tokenisation.[6] As our text data are in Chinese, we deploy the multilingual BERT, trained using word pieces from 100+ lexica, to derive textual information. On top of the BERT backbone, we add an MLP network for the downstream binary classification task and optimise the fine-tuned BERT model by selecting the best combination of learning rate and batch size in the search space of [4e-5, 3e-5, 2e-5, 1e-5, 1e-4] and [16, 32]. Further to this, for the combined BERT model, we concatenate the 768-dimensional pooled BERT representation with the structured data as input to the MLP layer, which we train together with BERT. An illustration of our combined BERT model architecture is shown in Figure 3.

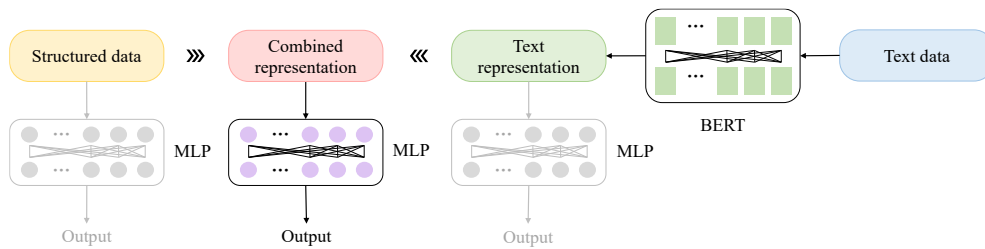

**Figure 3** Combined BERT model architecture

---

[6] To prevent BERT truncation for ChatGPT-refined texts, researchers could choose to set a maximum word count, guiding ChatGPT to produce text within this specified limit.

*4.2 Model training*

We divide our dataset into three distinct subsets – training, validation, and test – following the approach taken by Kriebel & Stitz (2022). The training set is used to train the models, the validation set is used to choose optimal model hyperparameters, and the test set is used to estimate out-of-sample prediction results. To construct these sets, we use the stratified random sampling method to select 70% of the samples as the training set, 20% of the training set as the validation set, and the remaining 30% as the test set. Thus, we have 1,377 observations in the training set, 345 in the validation set, and 738 in the test set.

Given the nature of our binary-label classification task, we chose the binary cross entropy (BCE) loss as the primary loss function for our MLP models. The BCE loss for the entire set of observations is denoted by $L(y, \hat{y})$, where $N$ represents the batch size, $y_i$ is the actual label (0 for non-defaulters and 1 for defaulters), and $\hat{y}_i$ is the predicted probability of default for observation $i$. We run training for multiple epochs and select hyperparameters that minimise the BCE loss on the validation set. Models with these optimised hyperparameters are saved and utilised to predict the probabilities of default for the test set.

$$L(y, \hat{y}) = -\frac{1}{N}\sum_{i=1}^{N}(y_i \cdot \log(\hat{y}_i) + (1 - y_i) \cdot \log(1 - \hat{y}_i)) \tag{1}$$

*4.3 Evaluation metrics*

Discrimination performance refers to the ability to distinguish between negative and positive classes. While there are several measures for gauging discrimination performance, it has been shown that comprehensively using multiple measures is the more reasonable approach (Lessmann et al., 2015). In this study, we consider four widely used metrics to evaluate the prediction results of our credit scoring models: the area under the receiver operating characteristic (ROC) curve (AUC), the Kolmogorov-Smirnov (KS) statistic, the H-measure, and the area under the precision recall (PR) curve (PRAUC). The ROC curve is a graphical representation of a classification model's predictive accuracy over a range of threshold values. The curve plots the true positive rate (i.e., sensitivity) against the false positive rate (i.e. 1 – specificity) for each threshold value of a given model (Wu & Li, 2021). The KS statistic is defined as the maximum vertical distance between the empirical cumulative distribution functions of the false positive rate and the true positive rate and can be used to assess the correctness of categorical predictions. Proposed by Hand (2009), the H-measure avoids the deficiency of the AUC of using different misclassification cost distributions for different classifiers, by specifying a preset severity ratio for assessing the impact of the misclassification cost between negative and positive instances. In this study, we adopt the standard severity ratio, set as the inverse of the relative class frequency (Chen et al., 2024). The PR curve has been cited as an alternative to the ROC curve for tasks with a large skew in the class distribution, which focuses on the trade-off between precision (i.e., the proportion of correctly classified positive instances among all predicted positives) and recall (i.e., the proportion of correctly classified positive instances among all actual positives) (Davis & Goadrich, 2006). Typically, the larger the AUC, KS, H-measure, and PRAUC, the better the performance of a prediction model.

To estimate the discrimination performance of each model, we initially employ five random seeds to conduct multiple runs of all experiments. Furthermore, we apply the bootstrap resampling method, conducting 1,000 resamples across each of the five random seeds (Berg-Kirkpatrick et al., 2012; Deldjoo, 2023; Katsafados et al., 2024). This approach yields a total of 5,000 performance estimates for the test set, thereby reducing randomness and enhancing the reliability and stability of the results. The discrimination performance results (mean and its 95% confidence interval) reported later are all based on the 5,000 estimates.

*4.4 Model interpretability*

In the business context, the interpretation of a model can play an essential role in data-driven decision-making. However, a deep learning model is often considered to be a black-box and not always well understood due to its complexity. In our study, to provide a better understanding of how textual input influences the prediction results of credit scoring models, we conduct content analysis using LIME. LIME is a model-agnostic technique that can be applied to any black-box model so that model predictions can be comprehended through the perturbing of the input data (Ribeiro et al., 2016). It is implemented as a surrogate model at a local level, assuming a linear relationship between the model's input and output, despite any non-linearity in the global relationship. It alters a single data sample by adjusting the feature values, and we can observe the effect on the output. When applied to the content analysis in this study, LIME produces new text by randomly removing words/phrases from the original text input, with each word/phrase represented as a binary feature (1 for included and 0 for removed).[7]

---

[7] SHAP (SHapley Additive exPlanations) is an alternative approach to exploring text importance. However, given that its usage would entail high computational costs for deep learning models and would ignore the direction of the relationship between the input variables and the response (Stevenson et al., 2021), we do not utilise it in our study.

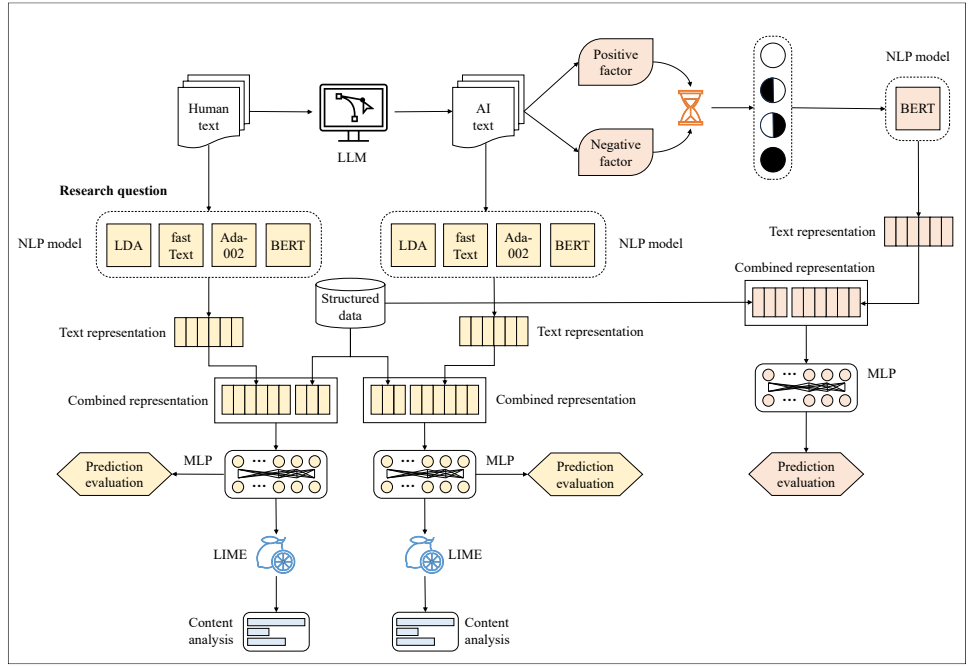

**Figure 4** Our empirical framework for default prediction

## 5. Empirical results

Figure 4 outlines our proposed research framework, illustrating how our research questions are addressed. Details and results are discussed in Sections 5.1 - 5.3, respectively. We also conduct further analysis in Section 5.4, which is introduced later.

*5.1 Prediction results of human-written and ChatGPT-refined texts*

In this section, we seek to examine the efficacy of text data for predicting defaults and compare the prediction results of human-written texts with those of ChatGPT-refined texts. Table 2 presents the discrimination performance of MLP models with the structured-only features, text-only features, and their combination, respectively. [8]

**Table 2** Model performance of human-written and ChatGPT-refined texts

| Model | Category | Structured | Text | Combined |
|---|---|---|---|---|
| | | | AUC | |
| LDA+MLP | Human | 0.616 (0.613, 0.618) | 0.572 (0.570, 0.574) | 0.667 (0.665, 0.669) |
| | ChatGPT | 0.616 (0.613, 0.618) | 0.651 (0.649, 0.653) | 0.687 (0.686, 0.689) |
| fastText+MLP | Human | 0.616 (0.613, 0.618) | 0.544 (0.542, 0.547) | 0.659 (0.657, 0.660) |
| | ChatGPT | 0.616 (0.613, 0.618) | 0.629 (0.626, 0.631) | 0.692 (0.690, 0.694) |
| Ada-002+MLP | Human | 0.616 (0.613, 0.618) | 0.618 (0.616, 0.621) | 0.689 (0.687, 0.691) |
| | ChatGPT | 0.616 (0.613, 0.618) | **0.670 (0.667, 0.672)** | 0.695 (0.693, 0.697) |
| BERT+MLP | Human | 0.616 (0.613, 0.618) | 0.629 (0.627, 0.632) | 0.667 (0.665, 0.670) |

---

[8] Our AUC scores are comparable to those reported in other customer loan default risk prediction studies, including Wang et al. (2020), Xia et al. (2020), Fitzpatrick & Mues (2021), Kriebel & Stitz (2022), Yu et al. (2023), Sanz-Guerrero & Arroyo (2024), and Babaei & Giudici (2024). These studies typically report the scores in the range of 0.60 to 0.75.

|  |  |  |  |  |
|---|---|---|---|---|
|  | ChatGPT | 0.616 (0.613, 0.618) | 0.660 (0.659, 0.662) | **0.710 (0.707, 0.712)** |
|  |  | KS | | |
| LDA+MLP | Human | 0.308 (0.305, 0.311) | 0.252 (0.250, 0.254) | 0.355 (0.353, 0.358) |
|  | ChatGPT | 0.308 (0.305, 0.311) | 0.359 (0.357, 0.362) | 0.388 (0.385, 0.390) |
| fastText+MLP | Human | 0.308 (0.305, 0.311) | 0.232 (0.229, 0.235) | 0.349 (0.347, 0.351) |
|  | ChatGPT | 0.308 (0.305, 0.311) | 0.325 (0.322, 0.328) | 0.390 (0.387, 0.393) |
| Ada-002+MLP | Human | 0.308 (0.305, 0.311) | 0.361 (0.357, 0.364) | 0.389 (0.386, 0.391) |
|  | ChatGPT | 0.308 (0.305, 0.311) | **0.409 (0.406, 0.413)** | 0.388 (0.385, 0.390) |
| BERT+MLP | Human | 0.308 (0.305, 0.311) | 0.322 (0.319, 0.325) | 0.392 (0.389, 0.395) |
|  | ChatGPT | 0.308 (0.305, 0.311) | 0.362 (0.359, 0.365) | **0.453 (0.450, 0.456)** |
|  |  | H-measure | | |
| LDA+MLP | Human | 0.186 (0.184, 0.189) | 0.180 (0.178, 0.182) | 0.271 (0.269, 0.274) |
|  | ChatGPT | 0.186 (0.184, 0.189) | 0.158 (0.156, 0.161) | 0.269 (0.266, 0.271) |
| fastText+MLP | Human | 0.186 (0.184, 0.189) | 0.165 (0.163, 0.168) | 0.266 (0.263, 0.268) |
|  | ChatGPT | 0.186 (0.184, 0.189) | 0.202 (0.199, 0.205) | 0.282 (0.279, 0.284) |
| Ada-002+MLP | Human | 0.186 (0.184, 0.189) | 0.267 (0.264, 0.270) | 0.267 (0.265, 0.269) |
|  | ChatGPT | 0.186 (0.184, 0.189) | **0.304 (0.301, 0.307)** | 0.275 (0.273, 0.278) |
| BERT+MLP | Human | 0.186 (0.184, 0.189) | 0.221 (0.218, 0.223) | 0.270 (0.267, 0.272) |
|  | ChatGPT | 0.186 (0.184, 0.189) | 0.201 (0.199, 0.203) | **0.294 (0.291, 0.297)** |
|  |  | PRAUC | | |
| LDA+MLP | Human | 0.086 (0.085, 0.088) | 0.195 (0.192, 0.197) | 0.157 (0.154, 0.159) |
|  | ChatGPT | 0.086 (0.085, 0.088) | 0.042 (0.041, 0.042) | 0.096 (0.095, 0.097) |
| fastText+MLP | Human | 0.086 (0.085, 0.088) | 0.190 (0.188, 0.193) | 0.231 (0.228, 0.233) |
|  | ChatGPT | 0.086 (0.085, 0.088) | 0.083 (0.082, 0.085) | 0.126 (0.124, 0.128) |
| Ada-002+MLP | Human | 0.086 (0.085, 0.088) | **0.237 (0.234, 0.240)** | **0.238 (0.235, 0.240)** |
|  | ChatGPT | 0.086 (0.085, 0.088) | 0.151 (0.149, 0.153) | 0.118 (0.116, 0.120) |
| BERT+MLP | Human | 0.086 (0.085, 0.088) | 0.205 (0.203, 0.208) | 0.203 (0.200, 0.206) |
|  | ChatGPT | 0.086 (0.085, 0.088) | 0.083 (0.081, 0.084) | 0.099 (0.098, 0.101) |

**Notes:** This table presents the discrimination performance of the AUC, KS, H-measure, and PRAUC (mean and its 95% confidence interval) for predicting the default risk of borrowers based on four NLP approaches: LDA+MLP, fastText+MLP, Ada-002+MLP, and BERT+MLP.

The average AUC, KS, H-measure, and PRAUC for the models relying solely on structured data are 0.616, 0.308, 0.186, and 0.086, respectively. As for the models trained on the text-only features, the majority of the prediction results when using the LDA+MLP and fastText+MLP models are worse than those generated by the structured-only models. In contrast, the text-only Ada-002+MLP and BERT+MLP models notably outperform the structured-only models, with Ada-002+MLP achieving the best average AUC, KS, H-measure, and PRAUC. These results highlight the success of the NLP techniques and suggest that contextualised models (e.g., Ada-002 and BERT) are more effective in capturing textual information than bag-of-words (LDA) and static word embedding (fastText) approaches. When we examine the performance of the combined models, as expected, we find that most of them outperform both the structured-only and text-only models and achieve impressive AUC, KS, H-measure, and PRAUC. These findings suggest that incorporating unstructured text data into credit scoring models can significantly improve prediction performance.

We further compare the models' predictive power using these two types of texts to identify default risk. Table 2 shows that, in most cases, models which use ChatGPT-refined texts exhibit far stronger results than those using human-written texts when using AUC, KS, and H-measure. Take the combined BERT+MLP model as an example: the average AUC, KS, and H-measure increase by 0.043, 0.061, and 0.024, respectively, when using ChatGPT-refined texts as opposed to human-written texts. Interestingly, the PRAUC results are not fully consistent with these findings, which suggest that ChatGPT-refined assessments are less effective at predicting positive cases (defaulters) compared to human-written assessments. For example, considering the text-only BERT+MLP models based on human-written and ChatGPT-refined texts, we find that the former achieves a higher PRAUC. Overall, these findings suggest that ChatGPT-refined texts enhance the predictive power of credit scoring models in terms of AUC, KS, and H-measure, but their impact on PRAUC remains less effective.[9,10] We further investigate the reason for this disagreement in Appendix D of the supplementary materials. Given these findings, we argue that while ChatGPT-refined texts may provide more insights than human-written texts by reducing noise, highlighting the most relevant and critical cues, and uncovering hidden information, they may also soften the language use and, in doing so, diminish the natural variability in expression between high-risk and low-risk borrowers. This could lead to a reduction in linguistic diversity, potentially masking subtle textual patterns that are informative for identifying credit risk. As robustness checks, we further use four alternative ML classifiers, namely LR, DT, RF and SVM, to supplement our main experiments; the related results, supporting our main findings, are available in Appendix G of the supplementary materials.

*5.2 Text content analysis*

In Section 5.1, our results indicate that text features extracted from loan assessments can serve as a powerful complement to traditional structured data, and their inclusion can significantly enhance prediction results when attempting to identify probabilities of default. In this section, we further explore which specific words/phrases can significantly influence the performance of credit scoring models. Given that the combined BERT+MLP model achieves superior performance in most cases, the following analysis will focus on this model specification to elucidate the influence of text content on model performance. Following Stevenson et al. (2021), we focus on uncertain cases from the test set, identified as those where the structured-only model yields predicted probabilities between 0.40 and 0.60, while an improvement in the predicted probabilities is observed when using the combined BERT+MLP model. Of this subset, for our analysis, we review 250 cases where we can see the greatest improvement in the combined model.

---

[9] We further examine the combined predictive power of human-written and ChatGPT-refined texts in predicting defaults by concatenating their derived text representations. The results presented in Appendix E of the supplementary materials reveal that the model using the combined text representations consistently outperforms those using only one type of text in most cases. This finding suggests that the information embedded within ChatGPT-refined text complements the content of the human-written text, highlighting the potential of generative AI in improving default prediction.

[10] To provide a more comprehensive and robust assessment of predictive model performance, we also compute additional evaluation metrics, including recall, precision, and F1 score (Maarouf et al., 2025), across various rejection thresholds - that is, by rejecting 70, 100, 120, 150, and 165 borrowers based on their predicted probability of default. The results are presented in Appendix F of the supplementary materials. The results for using these realistic rejection thresholds consistently suggest that ChatGPT-refined models outperform human-written models in terms of recall, precision, and F1 score.

In this analysis, we first employ LIME[11] to identify which specific words contribute most to the model's default prediction. Figures 5(a) and 5(b) present the top fifteen words derived from human-written and ChatGPT-refined texts, respectively, in terms of their influence on the predicted probabilities of default. The length of the bar represents the mean influence of the word on the model's prediction. Bars with positive values (in red) indicate that the inclusion of the word leads to an increase in the predicted probability of default, while bars with negative values (in green) denote a reduced predicted probability of default.

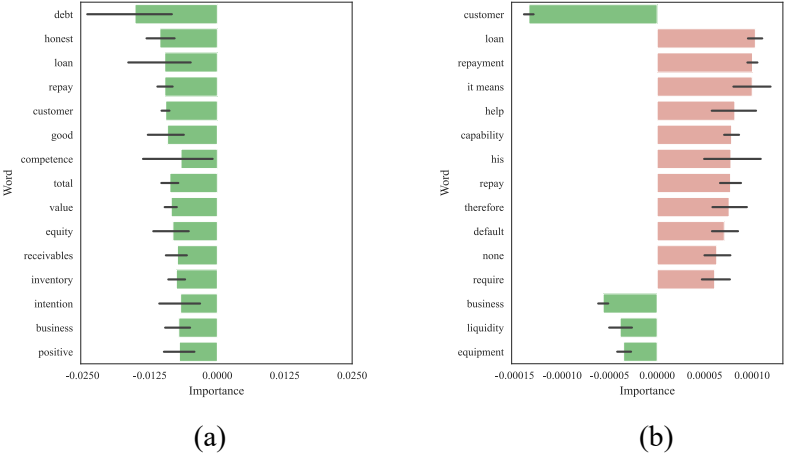

(a)        (b)

**Figure 5** Top 15 important words for: (a) human-written text, (b) ChatGPT-refined text

For those words sourced from human-written texts, we find that all of the top fifteen words shown in Figure 5(a) contribute to a reduction in the predicted probability of default. This suggests that these words are associated with a heightened perception of creditworthiness. Several themes emerge, including business operations, loan requirements, borrower characteristics, and words with a positive semantic orientation. The prevalence of certain words, such as 'customer', 'competence', and 'business', suggests that the model places significant emphasis on the borrower's capacity to manage the business. This, therefore, implies that a stable operational status is correlated with a reduced predicted probability of default. The model also associates a low default risk with terms such as 'equity', 'receivables', and 'inventory', words that describe the borrower's asset profile, which aligns with the common-sense view of loan review mechanisms. Furthermore, words pertaining to the borrower's loan requirements themselves, including 'debt', 'loan', and 'intention', may reflect the borrower's desire to obtain a loan for business development or to alleviate cash flow pressures and thus have a lower associated default prediction. The word 'repay' also reduces the predicted likelihood of default, indicating that repayment behaviour is an essential factor in the model's overall assessment. Some other words describing the personality traits of the borrower are strongly associated with the model's prediction of default risk. For instance, 'honest' and 'good' are explicit endorsements of the borrower and correspond to lower default prediction. The word 'positive' is another way the loan officers describe their assessments of a borrower's attitude and can reflect the borrower's willingness to provide documents and cooperate with the loan investigation process.

---

[11] The use of LIME offers managerial insights by identifying significant words and phrases that assist lenders in shaping future loan officer text descriptions. Stevenson et al. (2021) highlighted the potential of LIME to 'structure the unstructured data', which can enable lenders to capture more granular personal characteristics of borrowers than with standard model features. The analysis of text content can, therefore, provide a means to enhance loan assessments and offer valuable insights into the factors driving the credit risk assessment process.

For the ChatGPT-refined texts, the fifteen most important words can be categorised into themes related to business operations, loan requirements, and other words that lack ostensible factual meaning. In contrast to the results of the human-written texts, Figure 5(b) shows that the majority of these fifteen words show positive importance values, suggesting they are associated with higher predicted probabilities of default. Words such as 'customer', 'business', 'equipment', and 'liquidity' indicate that a borrower's stable operations and production can lead to decreased default predictions. In contrast to the roles loan requirement terms play in human-written texts, these terms relate to a higher predicted probability of default in almost all instances in the ChatGPT-refined texts. Loan requirement words such as 'loan', 'require', and 'help' convey the borrower's eagerness to relieve financial pressure by obtaining a loan and thus suggest a higher predicted probability of default. Some other loan repayment words, such as 'repayment', 'default', and 'none', also lead to a higher default risk. Moreover, some words that have no concrete meaning are nevertheless vital in affecting the model's predictions, such as 'it means' and 'his'. Although these words convey limited factual content, they notably affect the model's output, indicating that these words may often relate to other words that have significant influence on the model's predictions. Furthermore, the model associates a higher risk with the term 'capability', a finding which implies that, within the ChatGPT-refined texts, this term might frequently be employed to express scepticism regarding a borrower's capability to repay.

Interestingly, the words 'repay' and 'loan' themselves are associated with positive values when they appear in the ChatGPT-refined texts, suggesting that their inclusion can increase the predicted probability of default, which contrasts with the results derived from the human-written tests. These observations indicate that the importance of words for default risk prediction may be intricately tied to their contextualised embeddings. To further explore this possibility, we employ LIME to examine the importance of phrases or sentence fragments in both types of texts. In Chinese texts, punctuation marks serve as natural separators for sentence components, so we use them as delimiters to segment the text. This segmentation thereby facilitates a detailed analysis based on phrases or sentence fragments using LIME. For researchers interested in delving deeper, more advanced methods like *n*-gram algorithms or linguistic parsers could be considered to segment sentences before analysing them with LIME. These tools could analyse the syntactic structure of the text to identify phrases and sentence fragments more accurately. Figures 6(a) and 6(b) present the top fifteen phrases or sentence fragments that emerge as pivotal in the two types of texts.

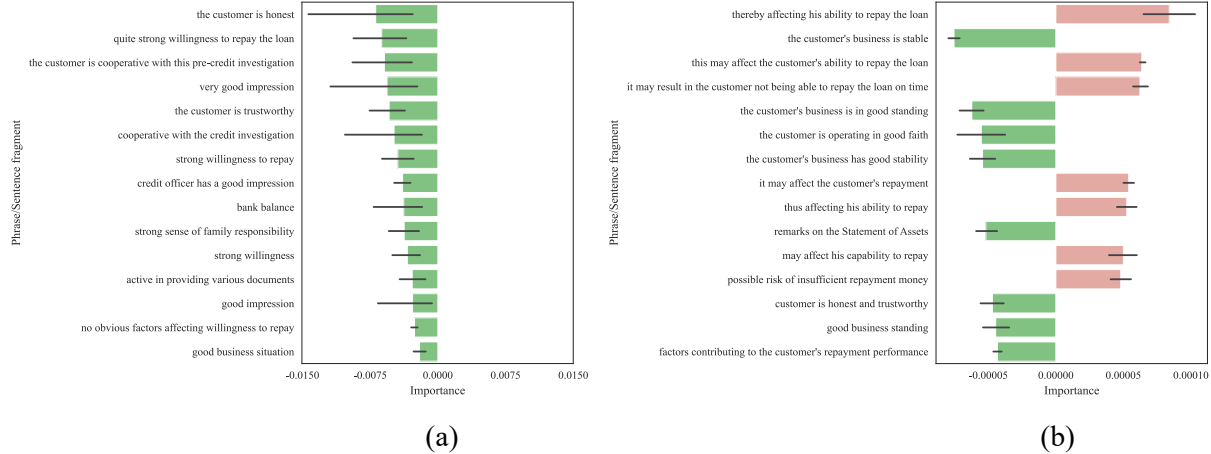

(a)            (b)

**Figure 6** Top 15 important phrases or sentence fragments for: (a) human-written text, (b) ChatGPT-refined text

Figure 6(a) shows that all of the fifteen most important phrases or sentence fragments are associated with negative values, suggesting that their inclusion in human-written text could reduce the model's predicted probability of default. This finding is consistent with our prior findings concerning word importance, as illustrated in Figure 5(a). The most important phrases/fragments are mainly related to the loan officer's comments on the borrower's credibility, reliability, and willingness to repay the loan, as observed during the pre-loan investigation. To illustrate, phrases such as 'strong willingness' and 'good impression' occur repeatedly in Figure 6(a). It is noteworthy that several words listed in Figure 5(a), such as 'customer', 'repay', and 'business', also appear in Figure 6(a). The contexts in which these words are embedded suggest a lowered predicted probability of default.

In contrast to the results for the human-written text, Figure 6(b) shows that only about half of the top fifteen phrases or sentence fragments from the ChatGPT-refined text have negative importance values. In line with the findings from the previous section, most of the negative phrases/fragments are related to the description of the borrower's business operations. For example, those such as 'the customer's business is stable', 'the customer's business is in good standing', and 'good business standing' all indicate healthy business operations, and therefore reduce the predicted probability of default. On the contrary, phrases or sentence fragments that increase the predicted risk of default are primarily associated with the borrower's compromised repayment capacity, such as 'thereby affecting his ability to repay the loan' and 'it may result in the customer not being able to repay the loan on time'. We also find that the words 'loan' and 'repay' are frequently present in phrases or fragments that convey risk, which is markedly different from their representations in human-written texts. Such insights highlight the notion that the valence – positive or negative – of terms in default prediction is profoundly influenced by their contextualised setting.

*5.3 Prediction results of positive and negative ChatGPT-refined factors*

In this section, we further explore the influence of different components of the ChatGPT-refined text in enhancing the model's prediction results, based on our best-performing model, BERT+MLP. As detailed in Section 3.3, we utilise ChatGPT to generate analytical assessments from human-written texts from two perspectives: positive factors supporting the borrower's repayment and negative factors that could potentially lead to the borrower's default. In this section, we examine how varying combinations of these factors affect model predictions. Table 3 presents the discrimination performance for five types of texts (i.e., human, positive factors only, negative factors only, positive followed by negative factors, and negative followed by positive factors). [12]

**Table 3** Model performance of different combinations of ChatGPT-refined factors

| Category | Structured | Text | Combined |
|---|---|---|---|
| | | AUC | |
| Human | 0.616 (0.613, 0.618) | 0.629 (0.627, 0.632) | 0.667 (0.665, 0.670) |
| Positive | 0.616 (0.613, 0.618) | 0.610 (0.607, 0.612) | 0.662 (0.659, 0.664) |
| Negative | 0.616 (0.613, 0.618) | 0.668 (0.666, 0.670) | 0.706 (0.703, 0.709) |
| Positive+Negative | 0.616 (0.613, 0.618) | 0.660 (0.659, 0.662) | 0.710 (0.707, 0.712) |
| Negative+Positive | 0.616 (0.613, 0.618) | **0.683 (0.681, 0.685)** | **0.719 (0.716, 0.721)** |
| | | KS | |
| Human | 0.308 (0.305, 0.311) | 0.322 (0.319, 0.325) | 0.392 (0.389, 0.395) |
| Positive | 0.308 (0.305, 0.311) | 0.304 (0.301, 0.308) | 0.378 (0.374, 0.382) |
| Negative | 0.308 (0.305, 0.311) | 0.385 (0.382, 0.388) | 0.449 (0.444, 0.453) |
| Positive+Negative | 0.308 (0.305, 0.311) | 0.362 (0.359, 0.365) | **0.453 (0.450, 0.456)** |
| Negative+Positive | 0.308 (0.305, 0.311) | **0.393 (0.390, 0.396)** | 0.449 (0.445, 0.453) |
| | | H-measure | |
| Human | 0.186 (0.184, 0.189) | 0.221 (0.218, 0.223) | 0.270 (0.267, 0.272) |
| Positive | 0.186 (0.184, 0.189) | 0.200 (0.197, 0.203) | 0.238 (0.235, 0.241) |
| Negative | 0.186 (0.184, 0.189) | 0.188 (0.185, 0.190) | 0.280 (0.276, 0.284) |
| Positive+Negative | 0.186 (0.184, 0.189) | 0.201 (0.199, 0.203) | **0.294 (0.291, 0.297)** |

---

[12] As previously discussed, BERT truncates texts exceeding 512 tokens. This truncation could lead to a potential loss of information essential for default prediction when handling texts longer than 512 tokens. Therefore, the positive+negative and negative+positive texts can generate different prediction outcomes.

| | | | |
|---|---|---|---|
| Negative+Positive | 0.186 (0.184, 0.189) | **0.233 (0.230, 0.236)** | 0.288 (0.284, 0.291) |
| | PRAUC | | |
| Human | 0.086 (0.085, 0.088) | **0.205 (0.203, 0.208)** | **0.203 (0.200, 0.206)** |
| Positive | 0.086 (0.085, 0.088) | 0.085 (0.083, 0.086) | 0.087 (0.086, 0.089) |
| Negative | 0.086 (0.085, 0.088) | 0.088 (0.087, 0.088) | 0.088 (0.087, 0.090) |
| Positive+Negative | 0.086 (0.085, 0.088) | 0.087 (0.085, 0.089) | 0.099 (0.098, 0.101) |
| Negative+Positive | 0.086 (0.085, 0.088) | 0.107 (0.105, 0.109) | 0.095 (0.094, 0.097) |

**Notes:** This table presents the discrimination performance of the AUC, KS, H-measure, and PRAUC (mean and its 95% confidence interval) of BERT+MLP for predicting the default risk of borrowers using original human-written text and different combinations of ChatGPT-refined factors.

Table 3 shows that, except for the case of positive factors only, all of the text-only models surpass the structured-only model. In alignment with our earlier findings, the combined models exhibit stronger performance than both the structured-only and text-only models in most cases. When considering AUC, KS, and H-measure, further analysis of different text types shows that models which utilise the negative+positive text generally outperform those using other text choices. Another noteworthy finding is that, once models incorporate the negative factors, they display stronger performance than those using the human-written text. However, the PRAUC results indicate that AI-refined text is less effective at predicting credit defaults compared to human-written texts. What is more, we observe that models focusing on the negative factors and the positive+negative text exhibit comparable performance on both the text-only and combined subsets. In light of these findings, we can conclude that incorporating comprehensive information derived from both positive and negative factors can produce superior results, with the negative factors identified by ChatGPT playing a crucial role in enhancing prediction performance. Lenders should therefore prioritise the understanding and evaluating of negative elements that might impact a borrower's capability to fulfil their payment obligations when it comes to gathering, processing, and analysing information.

*5.4 Further analysis: prediction results of different LLM-refined texts*

In this section, we conduct several additional experiments to validate our main findings illustrated in Section 5.1. In our main analysis, we rely on ChatGPT (version 2.13) to process and analyse the human-refined loan assessments. We now further investigate whether our main findings are sensitive to alternative generative AI models/versions, including ChatGPT4, ChatGLM2, and Baichuan2.[13] Table 4 presents the prediction results based on different LLM-refined texts.

**Table 4** Model performance of different LLM-refined texts

---

[13] ChatGPT4, building on OpenAI's GPT-4 model, has more parameters and training data than its predecessor, ChatGPT. The version of ChatGPT4 used in this paper is version 8.3, based on the GPT-4 model updated on 3rd August 2023. ChatGLM2 is the second-generation version of the open-source bilingual (Chinese-English) chat model ChatGLM proposed by Zhipu AI in 2022. More detailed information about ChatGLM2 can be found at https://huggingface.co/THUDM/chatglm2-6b. Baichuan2 is the new generation of large-scale open-source language models launched by Baichuan Intelligence inc. More detailed information about Baichuan2 can be found at https://huggingface.co/baichuan-inc/Baichuan2-13B-Chat.

| Model | Category | Structured | Text | Combined |
|---|---|---|---|---|
| | | | AUC | |
| LDA+MLP | Human | 0.616 (0.613, 0.618) | 0.572 (0.570, 0.574) | 0.667 (0.665, 0.669) |
| | ChatGPT4 | 0.616 (0.613, 0.618) | 0.591 (0.589, 0.593) | 0.668 (0.666, 0.670) |
| | ChatGLM2 | 0.616 (0.613, 0.618) | 0.633 (0.631, 0.634) | 0.665 (0.663, 0.667) |
| | Baichuan2 | 0.616 (0.613, 0.618) | 0.521 (0.518, 0.525) | 0.667 (0.665, 0.669) |
| fastText+MLP | Human | 0.616 (0.613, 0.618) | 0.544 (0.542, 0.547) | 0.659 (0.657, 0.660) |
| | ChatGPT4 | 0.616 (0.613, 0.618) | 0.617 (0.615, 0.619) | 0.664 (0.662, 0.666) |
| | ChatGLM2 | 0.616 (0.613, 0.618) | 0.632 (0.630, 0.634) | 0.665 (0.663, 0.667) |
| | Baichuan2 | 0.616 (0.613, 0.618) | 0.520 (0.518, 0.522) | 0.654 (0.652, 0.655) |
| Ada-002+MLP | Human | 0.616 (0.613, 0.618) | 0.618 (0.616, 0.621) | 0.689 (0.687, 0.691) |
| | ChatGPT4 | 0.616 (0.613, 0.618) | 0.681 (0.679, 0.682) | **0.733 (0.732, 0.735)** |
| | ChatGLM2 | 0.616 (0.613, 0.618) | 0.596 (0.593, 0.598) | 0.658 (0.656, 0.660) |
| | Baichuan2 | 0.616 (0.613, 0.618) | 0.624 (0.622, 0.626) | 0.674 (0.672, 0.676) |
| BERT+MLP | Human | 0.616 (0.613, 0.618) | 0.629 (0.627, 0.632) | 0.667 (0.665, 0.670) |
| | ChatGPT4 | 0.616 (0.613, 0.618) | **0.693 (0.691, 0.695)** | 0.718 (0.716, 0.720) |
| | ChatGLM2 | 0.616 (0.613, 0.618) | 0.618 (0.614, 0.621) | 0.633 (0.630, 0.636) |
| | Baichuan2 | 0.616 (0.613, 0.618) | 0.610 (0.608, 0.613) | 0.633 (0.631, 0.636) |
| | | | KS | |
| LDA+MLP | Human | 0.308 (0.305, 0.311) | 0.252 (0.250, 0.254) | 0.355 (0.353, 0.358) |
| | ChatGPT4 | 0.308 (0.305, 0.311) | 0.283 (0.280, 0.286) | 0.349 (0.347, 0.352) |
| | ChatGLM2 | 0.308 (0.305, 0.311) | **0.429 (0.427, 0.431)** | 0.346 (0.344, 0.349) |
| | Baichuan2 | 0.308 (0.305, 0.311) | 0.304 (0.301, 0.307) | 0.351 (0.349, 0.354) |
| fastText+MLP | Human | 0.308 (0.305, 0.311) | 0.232 (0.229, 0.235) | 0.349 (0.347, 0.351) |
| | ChatGPT4 | 0.308 (0.305, 0.311) | 0.289 (0.286, 0.291) | 0.338 (0.335, 0.340) |
| | ChatGLM2 | 0.308 (0.305, 0.311) | 0.348 (0.345, 0.351) | 0.350 (0.347, 0.353) |
| | Baichuan2 | 0.308 (0.305, 0.311) | 0.212 (0.210, 0.215) | 0.334 (0.331, 0.336) |
| Ada-002+MLP | Human | 0.308 (0.305, 0.311) | 0.361 (0.357, 0.364) | 0.389 (0.386, 0.391) |
| | ChatGPT4 | 0.308 (0.305, 0.311) | 0.378 (0.376, 0.380) | **0.463 (0.460, 0.465)** |
| | ChatGLM2 | 0.308 (0.305, 0.311) | 0.277 (0.274, 0.279) | 0.335 (0.333, 0.338) |
| | Baichuan2 | 0.308 (0.305, 0.311) | 0.333 (0.330, 0.337) | 0.356 (0.354, 0.359) |
| BERT+MLP | Human | 0.308 (0.305, 0.311) | 0.322 (0.319, 0.325) | 0.392 (0.389, 0.395) |
| | ChatGPT4 | 0.308 (0.305, 0.311) | 0.417 (0.415, 0.420) | 0.457 (0.454, 0.460) |
| | ChatGLM2 | 0.308 (0.305, 0.311) | 0.317 (0.313, 0.321) | 0.346 (0.342, 0.349) |
| | Baichuan2 | 0.308 (0.305, 0.311) | 0.304 (0.301, 0.307) | 0.347 (0.343, 0.350) |
| | | | H-measure | |
| LDA+MLP | Human | 0.186 (0.184, 0.189) | 0.180 (0.178, 0.182) | 0.271 (0.269, 0.274) |
| | ChatGPT4 | 0.186 (0.184, 0.189) | 0.114 (0.112, 0.116) | 0.263 (0.260, 0.265) |
| | ChatGLM2 | 0.186 (0.184, 0.189) | 0.179 (0.178, 0.181) | 0.260 (0.258, 0.263) |
| | Baichuan2 | 0.186 (0.184, 0.189) | 0.168 (0.165, 0.170) | 0.264 (0.262, 0.267) |
| fastText+MLP | Human | 0.186 (0.184, 0.189) | 0.165 (0.163, 0.168) | 0.266 (0.263, 0.268) |
| | ChatGPT4 | 0.186 (0.184, 0.189) | 0.190 (0.187, 0.192) | 0.249 (0.247, 0.252) |
| | ChatGLM2 | 0.186 (0.184, 0.189) | 0.218 (0.216, 0.221) | 0.258 (0.256, 0.261) |
| | Baichuan2 | 0.186 (0.184, 0.189) | 0.171 (0.169, 0.173) | 0.249 (0.247, 0.252) |
| Ada-002+MLP | Human | 0.186 (0.184, 0.189) | **0.267 (0.264, 0.270)** | 0.267 (0.265, 0.269) |
| | ChatGPT4 | 0.186 (0.184, 0.189) | 0.221 (0.219, 0.223) | **0.306 (0.304, 0.308)** |
| | ChatGLM2 | 0.186 (0.184, 0.189) | 0.209 (0.207, 0.212) | 0.201 (0.199, 0.203) |
| | Baichuan2 | 0.186 (0.184, 0.189) | 0.242 (0.239, 0.244) | 0.260 (0.257, 0.262) |
| BERT+MLP | Human | 0.186 (0.184, 0.189) | 0.221 (0.218, 0.223) | 0.270 (0.267, 0.272) |
|  | ChatGPT4 | 0.186 (0.184, 0.189) | 0.224 (0.222, 0.226) | 0.263 (0.260, 0.266) |
|  | ChatGLM2 | 0.186 (0.184, 0.189) | 0.188 (0.185, 0.192) | 0.201 (0.198, 0.203) |
|  | Baichuan2 | 0.186 (0.184, 0.189) | 0.168 (0.165, 0.170) | 0.201 (0.198, 0.204) |
|  |  | PRAUC | | |
| LDA+MLP | Human | 0.086 (0.085, 0.088) | 0.195 (0.192, 0.197) | 0.157 (0.154, 0.159) |
|  | ChatGPT4 | 0.086 (0.085, 0.088) | 0.097 (0.096, 0.097) | 0.114 (0.113, 0.116) |
|  | ChatGLM2 | 0.086 (0.085, 0.088) | 0.037 (0.036, 0.037) | 0.105 (0.104, 0.107) |
|  | Baichuan2 | 0.086 (0.085, 0.088) | 0.098 (0.096, 0.100) | 0.116 (0.114, 0.117) |
| fastText+MLP | Human | 0.086 (0.085, 0.088) | 0.190 (0.188, 0.193) | 0.231 (0.228, 0.233) |
|  | ChatGPT4 | 0.086 (0.085, 0.088) | 0.090 (0.088, 0.091) | 0.096 (0.094, 0.097) |
|  | ChatGLM2 | 0.086 (0.085, 0.088) | 0.073 (0.072, 0.074) | 0.105 (0.104, 0.106) |
|  | Baichuan2 | 0.086 (0.085, 0.088) | 0.090 (0.088, 0.093) | 0.093 (0.092, 0.094) |
| Ada-002+MLP | Human | 0.086 (0.085, 0.088) | 0.237 (0.234, 0.240) | **0.238 (0.235, 0.240)** |
|  | ChatGPT4 | 0.086 (0.085, 0.088) | 0.165 (0.163, 0.168) | 0.130 (0.129, 0.132) |
|  | ChatGLM2 | 0.086 (0.085, 0.088) | 0.159 (0.157, 0.162) | 0.063 (0.062, 0.064) |
|  | Baichuan2 | 0.086 (0.085, 0.088) | **0.242 (0.239, 0.245)** | 0.135 (0.133, 0.138) |
| BERT+MLP | Human | 0.086 (0.085, 0.088) | 0.205 (0.203, 0.208) | 0.203 (0.200, 0.206) |
|  | ChatGPT4 | 0.086 (0.085, 0.088) | 0.098 (0.097, 0.099) | 0.105 (0.103, 0.106) |
|  | ChatGLM2 | 0.086 (0.085, 0.088) | 0.092 (0.090, 0.094) | 0.067 (0.066, 0.068) |
|  | Baichuan2 | 0.086 (0.085, 0.088) | 0.098 (0.096, 0.100) | 0.067 (0.066, 0.068) |

**Notes:** This table presents the discrimination performance of the AUC, KS, H-measure, and PRAUC (mean and its 95% confidence interval) for predicting the default risk of borrowers based on four NLP approaches: LDA+MLP, fastText+MLP, Ada-002+MLP, and BERT+MLP across four types of texts, three of which are refined by LLMs.

In the table, we first observe that, in most instances, the combined models consistently show stronger performance than both the structured-only and text-only results. For example, the average AUC, KS, H-measure, and PRAUC increase by 0.102, 0.149, 0.077, and 0.019, respectively, for the combined BERT+MLP model built on the ChatGPT4-processed text, compared to the corresponding structured-only result. When we explore the capabilities of various generative AI models in analysing texts and the effectiveness of their refined texts in predicting default, we find some interesting results. Overall, ChatGPT4-refined texts generally exhibit superior performance across various NLP techniques and model subsets, particularly with the Ada-002+MLP and BERT+MLP models. The performance of the ChatGPT4-refined text is comparable with that of ChatGPT, which aligns with our expectations. Both ChatGPT and ChatGPT4 were developed by OpenAI and, given their similar modelling infrastructures, it is reasonable to expect a high degree of similarity in the content they generate (Feng et al., 2024). When comparing the prediction results of the ChatGLM2-refined and Baichuan2-refined texts, we find that their results are generally similar but vary depending on which NLP method is used. For example, for the Ada-002+MLP model, the Baichuan2 model achieves higher AUC, KS, H-measures, and PRAUC for both the text-only and combined subsets than the ChatGLM2 model. Conversely, the fastText+MLP models show superior performance when using the ChatGLM2-refined text. Moreover, the models relying on the ChatGLM2-refined and Baichuan2-refined texts do not demonstrate notable advantages over those based on the human-written texts when predicting default across all NLP approaches in terms of AUC, KS, and H-measure. However, similar with our prior results, we observe that ChatGPT-refined texts exhibit lower PRAUC scores compared to human-written texts. In summary, different generative AI models can differ significantly in their understanding, analysis, and response to a given text. Compared with other generative AI models, GPT-based models exhibit the most impressive performance in identifying borrowers' risks, thus enhancing overall default prediction results.

## 6. Discussion

In addition to the analysis in Section 5, this section further evaluates the business impact by estimating monetary gains associated with the models based on human-written and ChatGPT-refined assessments. Specifically, we follow the approach proposed by Kozodoi et al. (2025) to calculate the expected profit of our combined BERT+MLP model. For each loan, we consider the loan amount $L$ and interest rate $r$. In the event of a default (denoted as $D = 1$), the financial institution recovers $L \times (1 + r) \times (1 - LGD)$, where we assume a loss given default (LGD) of 0.9, reflecting the high cost of lending to defaulters. For a non-default borrower, the expected revenue is $L \times (1 + r)$. To assess overall profitability, we rank all borrowers in the test set in descending order based on their predicted probability of default and use equation 2 to calculate total profits when rejecting a certain number of borrowers.

$$profit = \sum_i^N [D_i \times L \times (1 + r) \times (1 - LGD) + (1 - D_i) \times L \times (1 + r) - L] \qquad (2)$$

Figure 7 illustrates the profit difference between using ChatGPT-refined and human-written assessments for rejection decisions in the combined BERT+MLP model. Positive values indicate that the ChatGPT-refined text model yields higher profits than the human-written text model, while negative values suggest the opposite. Since borrowers are rejected in descending order of predicted default probability, the $x$-axis represents a sequential exclusion of the highest-risk borrowers first, followed by medium-risk borrowers, and finally, low-risk borrowers.

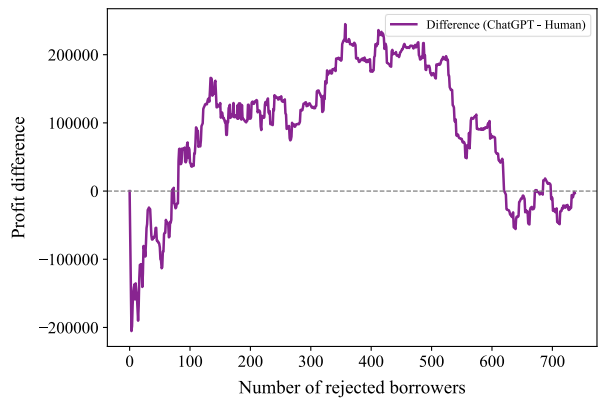

**Figure 7** Profit difference curve for the combined BERT+NLP model

The diagram shows that, with a small number of rejections (approximately the first 70), the human-written model appears to financially outperform the ChatGPT-refined text model. This underperformance may indicate that although ChatGPT can effectively extract relevant information from human-written loan assessments, uncover hidden insights, and infer additional cues for downstream tasks, it may also soften the language used, overgeneralise concerns related to the highest-risk borrowers, and highlight their potential positive factors in an effort to maintain consistency and generality in its responses. In contrast, human-written assessments often contain stronger or more implicit risk signals for the highest-risk borrowers. This may lead the credit scoring model based on ChatGPT-refined text to underestimate the highest-risk borrowers' default probabilities. Nevertheless, the negative profit difference diminishes rapidly as more borrowers are rejected. We observe a steady upward trend in profit difference, which turns positive after rejecting more than 70 borrowers and peaks at over ¥200,000 when approximately 350–380 borrowers are excluded. As lower-risk borrowers are subsequently rejected, the profit difference exhibits a downward trend but generally remains positive. Unlike the highest-risk borrowers, medium- and lower-risk borrowers in human-written assessments often have a mix of positive and negative information, along with some subjective noises, making their evaluations less straightforward. However, ChatGPT-refined texts tend to present information more clearly and consistently, which significantly helps the credit scoring model extract important information and reduce noise more effectively. In sum, these results suggest that, except in the highest-risk rejection range, ChatGPT-refined assessments enhance credit risk discrimination more effectively and profitably than human-written assessments.[14]

## 7. Conclusions

The rise of generative AI and advanced NLP techniques offers powerful new tools for extracting valuable insights from textual data, which can be instrumental in enhancing financial decision-making. Within the realm of credit scoring, several strategies have been adopted to incorporate LLMs to bolster predictions of borrower defaults. One common approach involves feeding both structured and textual borrowers' data into LLMs to directly assess credit risk levels, categorising borrowers as either high or low risk. Prior studies have demonstrated that, when a few training samples are provided in prompts, the prediction performance of LLMs can be improved significantly (Deldjoo, 2023; Babaei & Giudici, 2024). Nevertheless, LLMs used in this way often fall short of traditional credit scoring methods in their performance and struggle with the interpretability of results, particularly in the identification of key features influencing predictions. An alternative approach is to employ LLMs to convert textual data into numerical features, such as extracting psychological features from loan descriptions through the design of appropriate prompts (Yu et al., 2023). These text-derived features, integrated with structured features, can then be used in subsequent credit scoring models. This method's strength lies in its ability to quantify the impact of textual features on model predictions. However, its efficacy is heavily dependent on prompt content; variations in prompts can lead to significant numerical changes in the predictions and occasionally induce LLMs to generate unreliable hallucinations. Moreover, the potential applications of LLMs for credit scoring are vast, extending to the generation of synthetic examples for underrepresented classes to enhance small datasets, or the clustering and categorising of textual data to derive insightful analyses without the need for labelled training. Unlike previous studies, this paper explores a novel approach where LLMs serve as 'information intermediaries' that analyse and refine credit-related text, which is then processed by supervised NLP models to predict credit defaults.

---

[14] We also calculate recall, precision, and F1-score based on the threshold that maximises profit for each model, with the results presented in Appendix F of the supplementary materials. The results show that human-written models generally outperform ChatGPT-refined models across these metrics. This finding does not fully align with our results using other rejection thresholds. The main reason for this discrepancy is that when we choose to use the profit-maximising threshold, only a very small number of borrowers (fewer than 20) are rejected. As we discussed previously, compared to the human-written model, the ChatGPT-refined model seems less effective in identifying the highest-risk borrowers, leading to its underperformance in terms of recall, precision, and F1-score.

In our empirical experiments, we use ChatGPT to analyse the assessments produced by loan officers and thereby generate AI-refined loan assessments. Our findings reveal that there are significant differences between human-written and ChatGPT-refined texts in terms of text length, semantic similarity, and linguistic representations. When integrating unstructured text data into credit scoring models, the results indicate that both types of texts improve the performance of credit scoring models compared to those relying solely on structured data. While human-written texts seem to be more effective considering PRAUC, ChatGPT-refined texts achieve superior overall predictive performance in terms of AUC, KS, and H-measure. Furthermore, we explore what kinds of words/phrases significantly contribute to the predicted probability of default. The results show us that words related to business operations, loan requirements, and borrower characteristics provide more insights that influence a given model's predictions, and the importance of such words to these predictions is influenced significantly by their context. In addition, we examine which component of AI-refined texts has the most substantial influence on the models' performance enhancement. Our findings reveal that ChatGPT's analysis of borrower delinquency factors offers the most significant contribution. Moreover, we use several alternative generative AI tools for comparative analysis. These results further reinforce our conclusions and highlight the efficacy of GPT-based models in analysing textual information. We also estimate the monetary gains associated with models based on both human-written and ChatGPT-refined texts. The results indicate that the model relying on ChatGPT-refined texts yields higher profits than that using human-written texts, except in the highest-risk rejection range.

Our research not only contributes to the growing body of literature on the application of generative AI and NLP tools in the finance and business domain, but also has far-reaching managerial implications. First, as emerging LLMs are bringing about unprecedented changes over a wide range of industries (Chen et al., 2023a), this work highlights the importance of utilising LLMs and offers a new hands-on approach to using them in finance and business. Following the concepts of software-as-a-service, platform-as-a-service, and infrastructure-as-a-service, we believe that language-model-as-a-service will be the next core concept that could help enhance human exploration and understanding of the world around us (Zhao et al., 2023). Second, this study demonstrates that using generative AI tools to summarise and analyse human-written text can enhance lending decision-making, illustrating the strong capabilities of LLMs in both language comprehension and generation. Besides, the LLM-assisted problem-solving procedure we have introduced can easily be adopted by financial institutions for investment decision-making and risk management analysis. Third, as alternative data, such as images, video, audio, and text, become increasingly prevalent and more conveniently accessible (Chen et al., 2023b), our work serves as a valuable practical application of how to adequately leverage text data, especially AI-generated texts, to support finance and business-related decision-making procedures. These perspectives are of great significance to both research and practice.

This work inevitably has limitations that can be addressed in future studies. First, the content generated by LLMs can vary considerably due to the great flexibility of the prompts that can be used to guide them to generate output. Future studies could investigate how different prompt templates influence the results for the same research topic. Second, as OpenAI's ChatGPT and ChatGPT4 are not yet open-source, we faced challenges in conducting completely reproducible experiments, given that version updates may alter the results going forward. Furthermore, the generative AI tools we used are not tailored to the financial sector and analysis. Future research could explore using open-source LLMs and/or finance-specialised LLMs for default prediction. Third, the dataset size in our study is relatively small. Future research could benefit from employing larger datasets to further validate our findings. Moreover, the effectiveness of text data in credit default prediction may vary with loan types and durations. For instance, for long-term loans such as real estate loans, macroeconomic fluctuations and changes in a borrower's financial situation over time could affect the relevance of early text data for predicting credit default in subsequent years. Future research could explore how text data performs in predicting defaults across different loan types to deepen our understanding of these variations.

**Acknowledgements**

The authors would like to express their sincere gratitude to the associate editor and three anonymous reviewers for their constructive feedback, which helped to improve the quality of this paper significantly. We also thank Galina Andreeva, Anthony Bellotti, Cristián Bravo, Raffaella Calabrese, Jonathan Crook, Johannes Kriebel, Zhiyong Li, Zhao Wang, Huimin Zhao and conference and seminar participants at Credit Scoring & Credit Control Conference XVIII, Chinese Economics Association 2023 Annual Conference, University of Chinese Academy of Sciences, Dalian University of Technology, Dongbei University, Nanjing Agricultural University, and Yunnan University of Finance and Economics for their helpful feedback. This paper was supported by the Major Project of the National Social Science Foundation of China (No. 23&ZD175) and the National Natural Science Foundation of China (Nos. 71873103, 72173096).

# Supplementary materials

## Appendix A. Overview of structured features

**Table A.1** Definitions and statistical information of structured features

| Feature | Definition | Summary statistics | | | | |
|---|---|---|---|---|---|---|
| *Continuous* | | *Mean* | *SD* | *Min* | *Max* | *Originally missing (%)* |
| Loan amount ♣ | Loan amount (¥) | 90,550.81 | 22,839.84 | 1,000.00 | 200,000.00 | 0.00 |
| Loan rate ♣ | Loan interest rates (%) | 15.48 | 0.46 | 3.00 | 15.84 | 0.73 |
| Loan term ♣ | Loan term (months) | 3.07 | 0.47 | 1.00 | 8.00 | 0.00 |
| Rest amount | Remaining loan amount outstanding (¥) | 1,926.88 | 12,881.14 | 0.00 | 100,000.00 | 0.00 |
| Rest interest | Remaining loan interest outstanding (¥) | 88.17 | 734.32 | 0.00 | 11,322.52 | 0.00 |
| Rate-of-income | The ratio of monthly repayment to income (%) | 16.18 | 23.11 | 0.00 | 99.59 | 1.54 |
| Firm age ♣ | Number of years in operation | 5.50 | 4.30 | 0.00 | 36.00 | 0.28 |
| Space ♣ | Office space (m$^2$) | 1,025.67 | 3,094.97 | 0.00 | 33,300.00 | 1.38 |
| Employees ♣ | Number of employees | 11.50 | 26.56 | 0.00 | 500.00 | 0.45 |
| Procurement | Annual procurement frequency | 3.75 | 6.75 | 0.00 | 24.00 | 0.00 |
| Family members | Number of borrower's family members | 3.31 | 1.00 | 0.00 | 9.00 | 0.45 |
| Family workforce | Quantity of workforce available in borrower's family | 2.24 | 0.80 | 0.00 | 6.00 | 0.45 |
| Owner age ♣ | Age of business owner | 39.40 | 8.23 | 19.00 | 64.00 | 0.00 |
| *Categorical* | | *Number of categories* | | *Values* | | |
| Loan purpose ♣ | Types of loan purpose | 8 | | {1, 2, 3, 4, 5, 6, 7, 8} | | |
| Repay type | Types of loan repayment | 3 | | {1, 2, 3} | | |
| Advance payment | Types of early loan repayment | 2 | | {0, 1} | | |
| Business type ♣ | Types of borrowers' business | 4 | | {A, B, C, D} | | |
| Industry ♣ | Industry of borrower's business | 12 | | {A, D, E, F, G, H, I, J, K, M, N, P} | | |
| Customer type | Whether borrower is a new or existing customer of the bank | 2 | | {new_cust, old_cust} | | |
| Licence type | Types of business licence | 3 | | {0, 1, 2} | | |
| Credit rating | Borrower's credit rating | 8 | | {1, 2, 3, 4, 5, 6, 7, 8} | | |
| Ethnicity | Borrower's ethnicity | 10 | | {1, 2, 3, 4, 6, 7, 8, 9, 10, 11} | | |
| Education ♣ | Borrower's educational background | 9 | | {10, 20, 30, 40, 50, 60, 70, 80, 90} | | |
| Degree ♣ | Borrower's academic degree | 4 | | {0, 1, 2, 3} | | |

| | | | |
|---|---|---|---|
| Marital status ♣ | Borrower's marital status | 7 | {1, 2, 3, 4, 5, 6, 7} |
| Homeownership ♣ | Borrower's homeownership | 7 | {1, 2, 3, 4, 5, 6, 7} |
| Occupation ♣ | Borrower's occupation | 8 | {1, 2, 3, 4, 5, 6, Y, Z} |
| Job position ♣ | Borrower's job position | 5 | {A, B, C, D, E} |
| Job title ♣ | Borrower's job title | 5 | {0, 1, 2, 3, 4} |
| Postcode | Postcode of the business location | 19 | {200000, 214400, 628200, 215200, …}* |
| Verified ID ♣ | Borrower's citizenship category | 5 | {1, 2, 3, 4, 5} |

**Notes**: We conduct a data desensitisation process on the original structured data to protect private and business-sensitive information. * indicates that some infrequent categories are omitted, in the interest of space. ♣ indicates the final features selected for our credit scoring models.

**Appendix B. Examples of non-defaulters' human-written and ChatGPT-refined loan assessments**

Human-written text for a non-defaulter: *"The borrower is honest and has a clear business strategy. He was very cooperative with this loan investigation and actively provided various operational data. Historical data reflect the stability of his business operations. The borrower has been engaged in the metal building materials business for many years and has a wide range of connections. His main customers are well-acquainted friends. The borrower also has a strong willingness to repay. From our past investigations and feedback from the managers of the marketplace, the borrower's business is stable, and he has no bad habits. The credit system shows that the borrower has a good credit history and no overdue payment records. The borrower's previous loan payments were made on time, and he will renew the loan sooner. The balance of the borrower's credit card is ￥???. According to the unsettled delivery notes, the receivables are estimated to be ￥???, which basically matches the borrower's stated receivables of ￥???. The majority of the receivables will be settled within a month. We are conservatively estimating them to be ￥???. Additionally, the borrower has ￥??? of security deposit at a guarantee company. As physical inventory, the borrower has ??? tons of stock at a cost price of ￥??? and the estimated stock value is ￥???. We are cautiously recording this as an asset value of ￥???."*

ChatGPT-refined text for a non-defaulter: *"1. Factors supporting the borrower's repayment: * The borrower is trustworthy and has a clear business strategy. Historical information reflects his stable business operations and strong willingness to repay. * The borrower has been engaged in the metal building materials business for many years and has a wide range of connections. His main customers are well-acquainted friends, with whom he has a deep mutual understanding. * The credit system has revealed that the borrower has a good credit history with no record of overdue payments, and the borrower's previous repayments were very timely. 2. Factors that could lead to the borrower's default: * The balance on the borrower's bank card is ￥???. Financial strain might result in the borrower's inability to repay on time. * There are conservative estimates in the calculation of inventory value. If the actual value of the inventory is insufficient to cover the loan amount, it may also lead to the borrower's default."*

**Appendix C. Linguistic comparisons between human-written and ChatGPT-refined texts**

Figure C.1 illustrates that the average text length of the ChatGPT-refined texts is significantly longer than that of the original texts, their values being 355 and 209, respectively.[1] To evaluate the difference in average text lengths between the two, we conduct the Mann-Whitney U test (Mann & Whitney, 1947). This test yields a statistic of 654,395 with a *p*-value very close to 0, suggesting a significant difference at the 1% level between the two average text lengths. This notable disparity emerges because ChatGPT-refined texts generally offer more comprehensive and thorough assessments of a borrower's creditworthiness, thereby enhancing the richness of the information content.

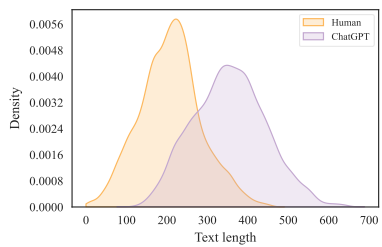

**Figure C.1** Text length of human-written and ChatGPT-refined texts

Furthermore, we calculate the cosine similarity between the embeddings of the two texts to evaluate the texts' semantic similarity. Cosine similarity is a vector-based measure used in a high-dimensional text representation space where similar texts are still close after projection (Tata & Patel, 2007). The cosine of the angle between two vectors reflects the semantic similarity of the corresponding texts. In this study, we employ three commonly used methods for text vectorisation: TF-IDF (Tata & Patel, 2007), GloVe (Pennington et al., 2014), and BERT (Devlin et al., 2018).[2] After obtaining the cosine similarity scores between human-written and ChatGPT-refined texts, we employ the Mann-Whitney U test to further examine whether the average cosine similarities of the two texts for the Bads and Goods groups are the same. The results shown in Table C.1 suggest a notable difference in the average text similarity across the different text vectorisation methods. Additionally, it is noteworthy that the cosine similarity scores for the Bads group are consistently lower than those for the Goods group. Figure C.2 visually confirms the above findings and shows that the Bads group exhibits a bimodal distribution, while the Goods group's distribution is more likely left-skewed. These results suggest that ChatGPT may possess some ability to presciently distinguish defaulters from non-defaulters, which may analyse the original textual loan assessments with a different focus.

---

[1] This result is also consistent with the findings of Guo et al. (2023), suggesting that ChatGPT tends to offer longer and more detailed responses.
[2] TF-IDF is a method that uses word frequencies to quantify the importance of word representations in a document taken from a collection of documents (also known as a corpus). GloVe is a pre-trained predictive algorithm for forming word embeddings that leverages global word-to-word co-occurrence counts based on the entire corpus. Unlike TF-IDF and GloVe, which ignore the word's contextual meaning, transformer-based BERT learns sequence-level semantics by considering the sequences of all words in the document, enabling it to identify different representations for words with more than one meaning.

**Table C.1** Cosine similarity statistics across Bads and Goods groups

| Model | Mean | | SD | | Statistic | *p*-value |
|---|---|---|---|---|---|---|
| | *Bads* | *Goods* | *Bads* | *Goods* | | |
| TF-IDF | 0.375 | 0.458 | 0.176 | 0.129 | 54,637 | 0.001*** |
| GloVe | 0.886 | 0.923 | 0.107 | 0.048 | 59,735 | 0.024** |
| BERT | 0.772 | 0.828 | 0.096 | 0.046 | 46,120 | 0.000*** |

**Notes**: *$p < 0.1$; **$p < 0.05$; ***$p < 0.01$.

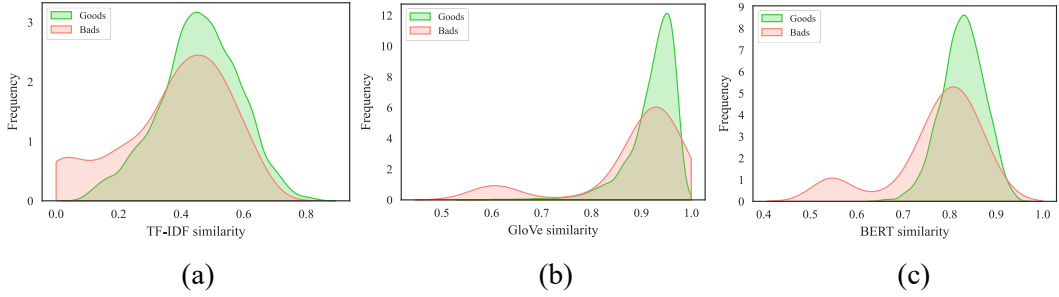

**Figure C.2** Cosine similarity distributions: (a) TF-IDF, (b) GloVe, (c) BERT

To further investigate the linguistic representations presented in the loan assessments generated by humans and ChatGPT, we use the linguistic inquiry and word count (LIWC) dictionary to assess the texts' relevance across various linguistic and psychological dimensions (Li et al., 2018). The LIWC dictionary, initially proposed by Tausczik & Pennebaker (2010), has been extensively employed in various domains, such as computational linguistics and social relationships. As our text data are in Chinese, we employ the Chinese LIWC dictionary (Huang et al., 2012) to compute the frequencies of words that belong to different semantic categories for both types of texts. Then, we use Welch's t-test to examine any semantic differences between the Bads and Goods groups in the two types of texts.[3] The results of Welch's t-tests for each linguistic, psychological, or topical category of the two types of texts for the Bads and Goods are presented in Tables C.2 and C.3, respectively.

**Table C.2** Welch's t-test results for the human-written and ChatGPT-refined texts for the Bads

| Category | Human | ChatGPT | t-stat | *p*-value | Category | Human | ChatGPT | t-stat | *p*-value |
|---|---|---|---|---|---|---|---|---|---|
| | *Affective words* | | | | Religion | 0.022 | 0.018 | -2.449 | 1.132 |
| Affect | 0.065 | 0.082 | 5.573 | 0.000*** | Work | 0.079 | 0.092 | 3.921 | 0.007** |
| Anger | 0.000 | 0.001 | 1.454 | 11.533 | | *Motive words* | | | |
| Anxiety | 0.001 | 0.006 | 9.176 | 0.000*** | Reward | 0.019 | 0.021 | 1.367 | 13.558 |
| Negations | 0.010 | 0.021 | 7.779 | 0.000*** | Risk | 0.008 | 0.015 | 5.823 | 0.000*** |

---

[3] Welch's t-statistic is calculated using the average values ($f_1$ and $f_2$) for a specific word category in the human-written and ChatGPT-refined texts, respectively. The total number of text tokens contributed by the human-written and ChatGPT-refined texts are denoted by $N_1$ and $N_2$. Following Li et al. (2018), we use the Bonferroni correction (M = 72) to adjust the significance level for each comparison. A positive t-statistic indicates an increased use of a semantic category in the ChatGPT-refined text compared to the original human-written text. In contrast, a negative t-statistic indicates a decreased use of a semantic category in the ChatGPT-refined text compared to the original human-written text. Welch's t-statistic is given by $t = \frac{f_2 - f_1}{\sqrt{\frac{f_1(1-f_1)}{N_1}} + \sqrt{\frac{f_2(1-f_2)}{N_2}}}$.

| | | | | | | | | | |
|---|---|---|---|---|---|---|---|---|---|
| Negative emotion | 0.010 | 0.029 | 13.209 | 0.000*** | | *Number words* | | | |
| Positive emotion | 0.042 | 0.040 | -0.812 | 32.945 | Numbers | 0.012 | 0.008 | -3.778 | 0.013* |
| Sadness | 0.001 | 0.002 | 3.372 | 0.059 | Quantities | 0.019 | 0.023 | 2.801 | 0.403 |
| Swear words | 0.002 | 0.000 | -2.904 | 0.292 | Quantity units | 0.121 | 0.097 | -6.161 | 0.000*** |
| | *Cognitive words* | | | | | *Perception words* | | | |
| Causation | 0.009 | 0.026 | 12.341 | 0.000*** | Feeling | 0.007 | 0.004 | -3.185 | 0.114 |
| Certainty | 0.014 | 0.016 | 1.187 | 18.595 | Hear | 0.002 | 0.007 | 7.786 | 0.000*** |
| Cognitive processes | 0.082 | 0.182 | 26.878 | 0.000*** | Motion | 0.021 | 0.031 | 5.869 | 0.000*** |
| Comparison | 0.027 | 0.034 | 3.103 | 0.152 | Perception | 0.031 | 0.027 | -2.350 | 1.482 |
| Differentiation | 0.007 | 0.025 | 12.997 | 0.000*** | See | 0.016 | 0.008 | -5.769 | 0.000*** |
| Discrepancy | 0.017 | 0.080 | 29.196 | 0.000*** | Space | 0.055 | 0.051 | -1.469 | 11.197 |
| Insight | 0.035 | 0.034 | -0.310 | 59.759 | | *Physical words* | | | |
| Relativity | 0.111 | 0.115 | 0.988 | 25.544 | Body | 0.009 | 0.009 | -0.498 | 48.855 |
| Tendentiousness | 0.031 | 0.041 | 4.509 | 0.001*** | Death | 0.001 | 0.001 | 0.593 | 43.705 |
| Tentative | 0.007 | 0.030 | 16.658 | 0.000*** | Food | 0.007 | 0.008 | 1.050 | 23.194 |
| | *Conversational words* | | | | Health | 0.003 | 0.002 | -0.707 | 37.89 |
| Assent | 0.024 | 0.037 | 6.411 | 0.000*** | Physical | 0.020 | 0.019 | -0.413 | 53.672 |
| Fillers | 0.004 | 0.004 | -0.021 | 77.658 | Sexual | 0.002 | 0.002 | -0.614 | 42.591 |
| Netspeak | 0.017 | 0.025 | 4.998 | 0.000*** | | *Pronoun words* | | | |
| Nonfluencies | 0.000 | 0.001 | 5.295 | 0.000*** | 1st person plural | 0.001 | 0.000 | -3.616 | 0.024* |
| | *Psychological words* | | | | 1st person singular | 0.003 | 0.000 | -4.917 | 0.000*** |
| Achievement | 0.016 | 0.033 | 9.885 | 0.000*** | 3rd person plural | 0.000 | 0.001 | 5.661 | 0.000*** |
| Affiliation | 0.012 | 0.006 | -4.472 | 0.001*** | 3rd person singular | 0.001 | 0.004 | 6.197 | 0.000*** |
| Drives | 0.095 | 0.132 | 10.013 | 0.000*** | Impersonal pronouns | 0.008 | 0.004 | -4.600 | 0.000*** |
| Power | 0.046 | 0.075 | 10.796 | 0.000*** | Personal pronouns | 0.005 | 0.005 | 0.470 | 50.408 |
| | *Function words* | | | | Pronouns | 0.013 | 0.009 | -3.209 | 0.105 |
| Adverbs | 0.068 | 0.080 | 3.624 | 0.023* | | *Social words* | | | |
| Auxiliary verbs | 0.043 | 0.102 | 20.987 | 0.000*** | Family | 0.001 | 0.000 | -1.922 | 4.315 |
| Conjunctions | 0.020 | 0.051 | 15.747 | 0.000*** | Female references | 0.002 | 0.000 | -2.952 | 0.25 |
| Function words | 0.383 | 0.495 | 19.214 | 0.000*** | Friends | 0.000 | 0.000 | -0.170 | 68.343 |
| Interrogatives | 0.002 | 0.001 | -2.073 | 3.013 | Male references | 0.000 | 0.000 | -0.555 | 45.757 |
| Postpositions | 0.014 | 0.013 | -0.916 | 28.416 | Social processes | 0.053 | 0.057 | 1.619 | 8.327 |
| Prepositions | 0.056 | 0.057 | 0.473 | 50.267 | | *Time words* | | | |
| Programmes | 0.004 | 0.003 | -1.597 | 8.711 | Future focus | 0.008 | 0.024 | 12.209 | 0.000*** |
| | *Lifestyle words* | | | | Past focus | 0.006 | 0.004 | -1.961 | 3.944 |
| Home | 0.003 | 0.002 | -0.995 | 25.249 | Present focus | 0.015 | 0.011 | -2.637 | 0.662 |
| Leisure | 0.006 | 0.002 | -4.948 | 0.000*** | Time | 0.048 | 0.036 | -4.946 | 0.000*** |
| Money | 0.106 | 0.098 | -2.115 | 2.722 | | | | | |

**Notes**: 'Human' and 'ChatGPT' denote the original human-written and ChatGPT-refined textual loan assessment, respectively. *$p$ < 0.1/M; **$p$ < 0.05/M; ***$p$ < 0.01/M.

From Table C.2, we find that, for the Bads group, there are significant differences in word frequencies across 36 out of the 72 word categories when comparing the human-written and ChatGPT-refined texts. The LIWC dictionary's division of word categories shows that the most notable difference between the two types of text can be observed in words associated with cognitive words, pronouns, psychological words, and affective words. Specifically, ChatGPT-refined texts exhibit a higher usage of cognitive words, such as words in the 'discrepancy', 'cognitive processes', 'tentative', 'differentiation', 'causation', and 'tendentiousness' categories. This indicates ChatGPT's ability to logically analyse human-written texts and incorporate multiple perspectives. Furthermore, we observe that ChatGPT uses fewer first-person and informal pronouns, preferring more third-person pronouns than loan officers. This pattern may reflect ChatGPT's objective standpoint. Regarding psychological words, we find an increased usage of categories like 'drives', 'achievement', and 'power' in ChatGPT's output, while 'affiliation' words show a decreased usage. These findings suggest ChatGPT's proficiency in identifying the borrower's personality traits and the potential for non-default outcomes. Furthermore, we find ChatGPT tends to use more affective words, such as 'negative emotion', 'anxiety', 'negations', and 'affect' words, than the loan officers. Additionally, we observe notable differences in conversational words, perception words, and time words between the human-written and ChatGPT-refined texts.

**Table C.3** Welch's t-test results for the human-written and ChatGPT-refined texts for the Goods

| Category | Human | ChatGPT | t-stat | p-value | Category | Human | ChatGPT | t-stat | p-value |
|---|---|---|---|---|---|---|---|---|---|
| *Affective words* | | | | | Religion | 0.023 | 0.023 | 0.055 | 75.555 |
| Affect | 0.063 | 0.084 | 47.346 | 0.000*** | Work | 0.079 | 0.092 | 25.630 | 0.000*** |
| Anger | 0.001 | 0.001 | 8.325 | 0.000*** | *Motive words* | | | | |
| Anxiety | 0.001 | 0.006 | 60.811 | 0.000*** | Reward | 0.016 | 0.022 | 24.255 | 0.000*** |
| Negations | 0.014 | 0.020 | 30.622 | 0.000*** | Risk | 0.007 | 0.014 | 42.353 | 0.000*** |
| Negative emotion | 0.010 | 0.027 | 72.095 | 0.000*** | *Number words* | | | | |
| Positive emotion | 0.042 | 0.045 | 7.389 | 0.000*** | Numbers | 0.013 | 0.008 | -28.525 | 0.000*** |
| Sadness | 0.000 | 0.002 | 21.181 | 0.000*** | Quantities | 0.019 | 0.023 | 15.153 | 0.000*** |
| Swear words | 0.001 | 0.001 | -14.222 | 0.000*** | Quantity units | 0.128 | 0.099 | -49.532 | 0.000*** |
| *Cognitive words* | | | | | *Perception words* | | | | |
| Causation | 0.009 | 0.025 | 72.801 | 0.000*** | Feeling | 0.005 | 0.003 | -14.205 | 0.000*** |
| Certainty | 0.013 | 0.016 | 12.559 | 0.000*** | Hear | 0.003 | 0.006 | 33.665 | 0.000*** |
| Cognitive processes | 0.080 | 0.167 | 157.364 | 0.000*** | Motion | 0.022 | 0.028 | 25.483 | 0.000*** |
| Comparison | 0.030 | 0.031 | 5.590 | 0.000*** | Perception | 0.027 | 0.027 | -2.945 | 0.255 |
| Differentiation | 0.011 | 0.021 | 45.134 | 0.000*** | See | 0.015 | 0.009 | -30.416 | 0.000*** |
| Discrepancy | 0.016 | 0.071 | 168.183 | 0.000*** | Space | 0.057 | 0.049 | -19.073 | 0.000*** |
| Insight | 0.033 | 0.036 | 8.239 | 0.000*** | *Physical words* | | | | |
| Relativity | 0.111 | 0.109 | -4.478 | 0.001*** | Body | 0.009 | 0.008 | -5.572 | 0.000*** |
| Tendentiousness | 0.026 | 0.039 | 41.740 | 0.000*** | Death | 0.001 | 0.001 | 0.547 | 46.174 |
| Tentative | 0.005 | 0.023 | 92.908 | 0.000*** | Food | 0.006 | 0.006 | 0.644 | 41.028 |
| *Conversational words* | | | | | Health | 0.003 | 0.002 | - | 0.000*** |
| Assent | 0.025 | 0.038 | 43.343 | 0.000*** | Physical | 0.019 | 0.017 | 12.987 -8.785 | 0.000*** |
| Fillers | 0.004 | 0.004 | 0.696 | 38.407 | Sexual | 0.002 | 0.003 | 6.711 | 0.000*** |
| Netspeak | 0.020 | 0.023 | 12.323 | 0.000*** | *Pronoun words* | | | | |
| Nonfluencies | 0.000 | 0.000 | 1.386 | 13.102 | 1st person plural | 0.002 | 0.000 | -27.811 | 0.000*** |
| *Psychological words* | | | | | 1st person singular | 0.003 | 0.000 | -35.243 | 0.000*** |
| Achievement | 0.015 | 0.031 | 61.731 | 0.000*** | 3rd person plural | 0.000 | 0.000 | 9.735 | 0.000*** |
| Affiliation | 0.013 | 0.008 | -28.067 | 0.000*** | 3rd person singular | 0.001 | 0.002 | 20.139 | 0.000*** |
| Drives | 0.090 | 0.133 | 79.726 | 0.000*** | Impersonal pronouns | 0.008 | 0.004 | -27.482 | 0.000*** |
| Power | 0.044 | 0.074 | 74.774 | 0.000*** | Personal pronouns | 0.005 | 0.002 | -27.328 | 0.000*** |
| *Function words* | | | | | Pronouns | 0.013 | 0.006 | -38.650 | 0.000*** |
| Adverbs | 0.067 | 0.078 | 25.170 | 0.000*** | *Social words* | | | | |
| Auxiliary verbs | 0.043 | 0.098 | 128.758 | 0.000*** | Family | 0.002 | 0.001 | -11.758 | 0.000*** |
| Conjunctions | 0.016 | 0.046 | 105.479 | 0.000*** | Female references | 0.002 | 0.001 | -19.993 | 0.000*** |
| Function words | 0.380 | 0.475 | 108.784 | 0.000*** | Friends | 0.001 | 0.000 | -10.778 | 0.000*** |
| Interrogatives | 0.002 | 0.002 | -5.618 | 0.000*** | Male references | 0.000 | 0.000 | -6.836 | 0.000*** |
| Postpositions | 0.014 | 0.011 | -14.078 | 0.000*** | Social processes | 0.058 | 0.060 | 3.177 | 0.118 |
| Prepositions | 0.055 | 0.055 | 0.235 | 64.315 | *Time words* | | | | |
| Programmes | 0.004 | 0.004 | -5.441 | 0.000*** | Future focus | 0.007 | 0.023 | 78.707 | 0.000*** |
| *Lifestyle words* | | | | | Past focus | 0.006 | 0.004 | -12.476 | 0.000*** |
| Home | 0.004 | 0.003 | -12.590 | 0.000*** | Present focus | 0.010 | 0.010 | -3.519 | 0.034* |
| Leisure | 0.005 | 0.002 | -27.458 | 0.000*** | Time | 0.045 | 0.034 | -29.912 | 0.000*** |
| Money | 0.104 | 0.093 | -20.275 | 0.000*** | | | | | |

**Notes**: 'Human' and 'ChatGPT' denote the original human-written and ChatGPT-refined textual loan assessment, respectively. *$p < 0.1$/M; **$p < 0.05$/M; ***$p < 0.01$/M.

Table C.3 shows that the usage of 64 word categories differs significantly between the two types of texts for the Goods, which indicates that the semantic similarity between these two types of texts is relatively lower for non-defaulters than for defaulters. The differences are primarily observed in categories associated with cognitive, affective, pronoun, and perceptive words. Consistent with the results presented in Table C.2, the ChatGPT-refined texts exhibit an increased usage of cognitive words, including 'discrepancy', 'cognitive processes', 'tentative', 'causation', 'differentiation', 'tendentiousness', 'certainty', 'insight', and 'comparison' words. Moreover, the ChatGPT-refined texts tend to include more affective words, such as 'negative emotion', 'anxiety', 'affect', 'negations', 'sadness', 'anger', and 'positive emotion' words, than the human-written texts. Conversely, words related to perception, such as 'see', 'feeling', and 'space' words, are more frequently used in the human-written texts, while 'motion' words are more commonly used in the

ChatGPT-refined texts. Other categories, including function, physical, and social words, also exhibit significant differences between the human-written and ChatGPT-refined texts.

**Appendix D. Understanding the limited performance of PRAUC for ChatGPT-refined models**

To further investigate the potential reason behind the underperformance of PRAUC in models relying on the ChatGPT-refined text compared to the human-written text, we have conducted additional analyses and visualisations. Specifically, we have visualised the sorted default prediction scores with defaulters for both the text-only and combined BERT+MLP models using human-written and ChatGPT-refined texts (referred to as the Human model and ChatGPT model, respectively) in Figures D.1 and D.3. In addition, we have plotted their ROC and PR curves in Figures D.2 and D.4, respectively.

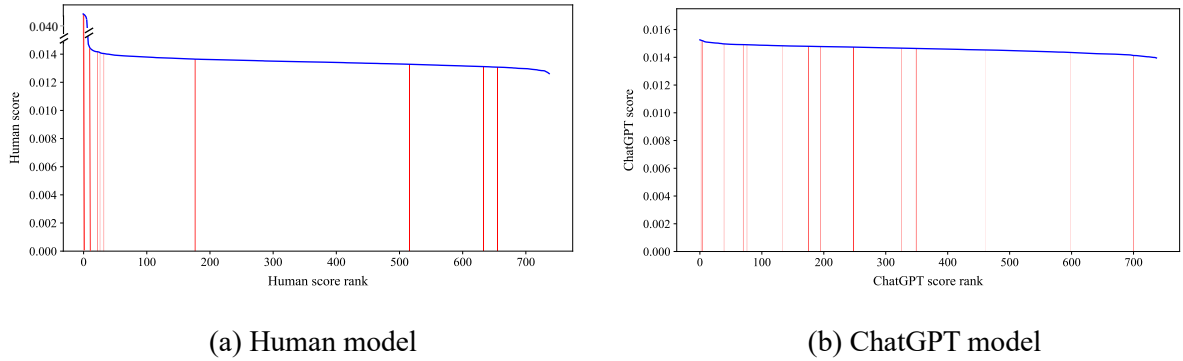

(a) Human model  (b) ChatGPT model

**Figure D.1** Text-only models' default prediction scores of (a) the Human model and (b) the ChatGPT model sorted in descending order (blue lines) corresponding to all borrowers in the test set. Red vertical lines denote defaulters. The average rank of the Human model's prediction scores for defaulters is 271, while the ChatGPT model's prediction scores for defaulters have a lower average rank of 244.

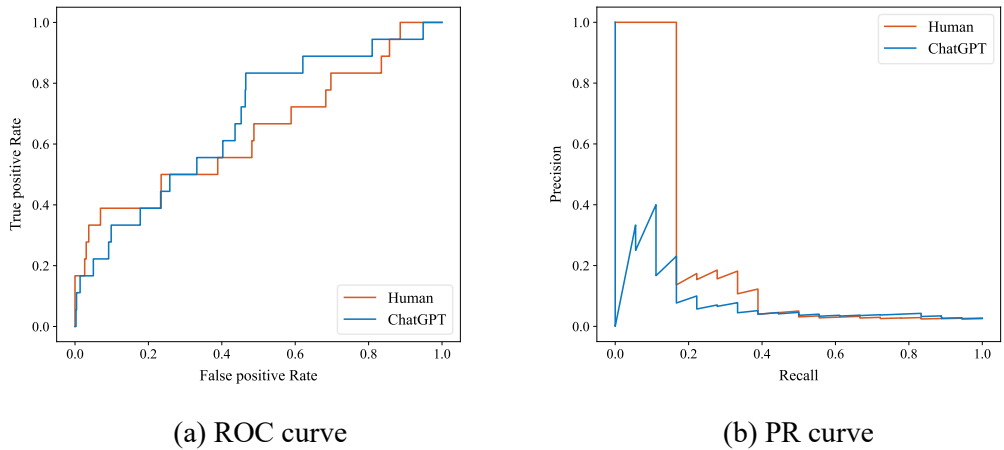

(a) ROC curve  (b) PR curve

**Figure D.2** Comparison of (a) ROC and (b) PR curves for text-only models' default prediction (test set).

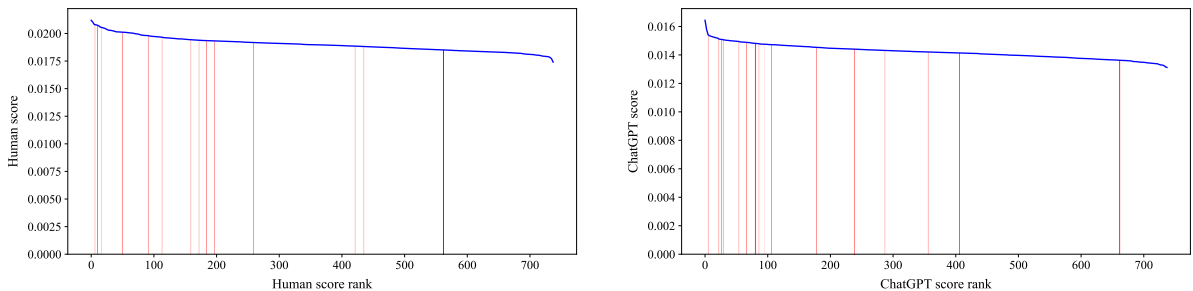

(a) Human model                                          (b) ChatGPT model

**Figure D.3** Combined models' default prediction scores of (a) the Human model and (b) the ChatGPT model sorted in descending order (blue lines) corresponding to all borrowers in the test set. Red vertical lines denote defaulters. The average rank of the Human model's prediction scores for defaulters is 237, while the ChatGPT model's prediction scores for defaulters have a lower average rank of 192.

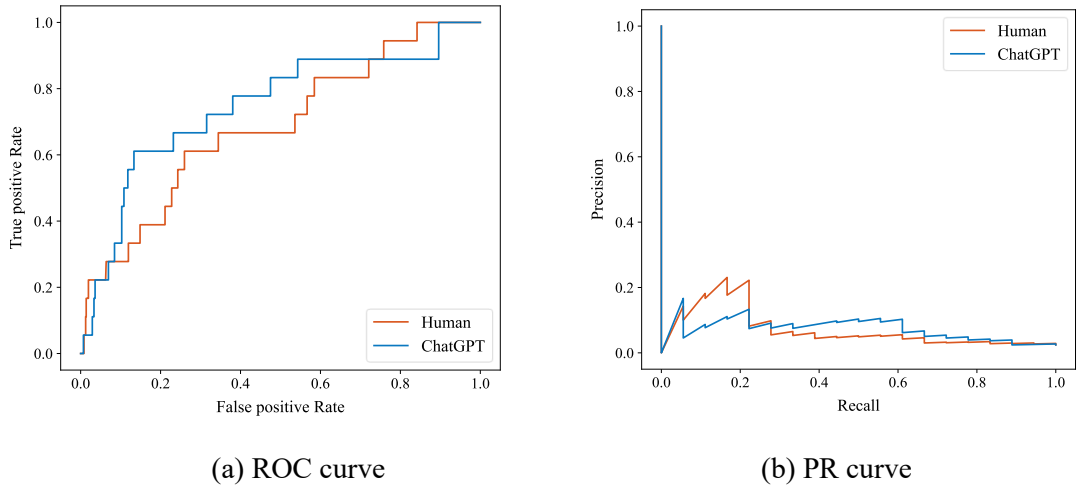

(a) ROC curve                                            (b) PR curve

**Figure D.4** Comparison of (a) ROC and (b) PR curves for combined models' default prediction (test set).

In Figures D.1 and D.3, we observe a consistent pattern where the Human models perform better at the highest scores but fail to identify many defaulters in the middle and lower score ranges. This results in higher precision and a lower false positive rate (FPR) for low values of recall, but lower precision and a higher FPR for high values of recall, presented in Figures D.2 and D.4. Consequently, this discrepancy leads to a disagreement between AUC and PRAUC. Since the PR curve primarily emphasises the accurate prediction of positive cases, the Human models are rewarded for their higher precision at high scores while being less penalised for missing defaulters at low scores (Davis & Goadrich, 2006). This bias towards the positive class suggests that PRAUC favours models with good performance in positive prediction, while AUC offers a more balanced consideration in both positive and negative predictions (Junuthula et al., 2016).

# Appendix E. Prediction results of combined human-written and ChatGPT-refined text representations

Table E.1 Model performance of combined human-written and ChatGPT-refined text representations

| Model | Category | Structured | Text | Combined |
|---|---|---|---|---|
| | | | AUC | |
| LDA+MLP | Human+ChatGPT | 0.616 (0.613, 0.618) | 0.646 (0.644, 0.648) | 0.695 (0.693, 0.697) |
| fastText+MLP | Human+ChatGPT | 0.616 (0.613, 0.618) | 0.640 (0.638, 0.643) | 0.688 (0.686, 0.690) |
| Ada-002+MLP | Human+ChatGPT | 0.616 (0.613, 0.618) | 0.635 (0.632, 0.637) | 0.707 (0.705, 0.709) |
| BERT+MLP | Human+ChatGPT | 0.616 (0.613, 0.618) | **0.674 (0.672, 0.676)** | **0.738 (0.736, 0.740)** |
| | | | KS | |
| LDA+MLP | Human+ChatGPT | 0.308 (0.305, 0.311) | 0.338 (0.335, 0.340) | 0.388 (0.385, 0.390) |
| fastText+MLP | Human+ChatGPT | 0.308 (0.305, 0.311) | 0.363 (0.360, 0.365) | 0.363 (0.361, 0.366) |
| Ada-002+MLP | Human+ChatGPT | 0.308 (0.305, 0.311) | **0.385 (0.382, 0.388)** | 0.418 (0.415, 0.420) |
| BERT+MLP | Human+ChatGPT | 0.308 (0.305, 0.311) | 0.365 (0.362, 0.367) | **0.469 (0.468, 0.471)** |
| | | | H-measure | |
| LDA+MLP | Human+ChatGPT | 0.186 (0.184, 0.189) | 0.221 (0.219, 0.224) | 0.282 (0.279, 0.284) |
| fastText+MLP | Human+ChatGPT | 0.186 (0.184, 0.189) | 0.246 (0.243, 0.248) | 0.273 (0.271, 0.276) |
| Ada-002+MLP | Human+ChatGPT | 0.186 (0.184, 0.189) | **0.303 (0.300, 0.306)** | 0.283 (0.280, 0.285) |
| BERT+MLP | Human+ChatGPT | 0.186 (0.184, 0.189) | 0.248 (0.246, 0.251) | **0.334 (0.332, 0.337)** |
| | | | PRAUC | |
| LDA+MLP | Human+ChatGPT | 0.086 (0.085, 0.088) | 0.207 (0.204, 0.210) | 0.171 (0.168, 0.173) |
| fastText+MLP | Human+ChatGPT | 0.086 (0.085, 0.088) | 0.216 (0.213, 0.218) | 0.245 (0.242, 0.248) |
| Ada-002+MLP | Human+ChatGPT | 0.086 (0.085, 0.088) | **0.252 (0.249, 0.255)** | 0.236 (0.233, 0.239) |
| BERT+MLP | Human+ChatGPT | 0.086 (0.085, 0.088) | 0.225 (0.222, 0.228) | **0.253 (0.251, 0.256)** |

Notes: This table presents the discrimination performance of AUC, KS, H-measure, and PRAUC (mean and its 95% confidence interval) for predicting the default risk of borrowers based on four NLP approaches: LDA+MLP, fastText+MLP, Ada-002+MLP, and BERT+MLP.

In Table E.1, we observe that using integrated text representations (Human+ChatGPT) can achieve the best prediction performance in most cases. For example, the combined BERT+MLP model achieves the highest AUC score of 0.738 when using integrated text representations, outperforming both ChatGPT-refined text only (0.710) and human-written text only (0.667). These results suggest that integrated text representations effectively capture the information contained in both human-written and ChatGPT-refined texts. Moreover, they indicate that insights gained from ChatGPT-refined text can complement those from human-written text, potentially boosting overall model predictive performance. These findings further reinforce the validity and practicality of our proposed framework.

**Appendix F. Robustness check across alternative metrics**

In this appendix, we have calculated the recall, precision, and F1 score based on the threshold that maximises profit for each model and using fixed thresholds by rejecting a specific number of borrowers (i.e., 70, 100, 120, 150, and 165 rejections). For the results based on the threshold that maximises profit shown in Table F.1, we find that human-written models generally outperform ChatGPT-refined models across these metrics. This finding does not fully align with our results using other rejection thresholds, which are presented shortly. The primary reason for this discrepancy is that when the profit-maximising threshold is applied, only a very small number of borrowers (fewer than 20) are rejected. As we discussed previously, compared to the human-written model, the ChatGPT-refined model seems less effective in identifying the highest-risk borrowers, leading to its underperformance in terms of recall, precision, and F1-score. These findings are also reflected in the profit difference curve shown in "Figure 7", where the ChatGPT-refined model is less profitable than the human-written model at very low rejection rates.

In contrast, the results for using other fixed rejection thresholds (i.e., ranging from 70 to 165 rejections) in Tables F.1 and F.2 show that ChatGPT-refined models outperform human-written models in most cases across these thresholds. As the number of rejected borrowers increases, recall steadily improves while precision remains relatively stable with only a slight decrease. These results indicate that ChatGPT-refined models not only enhance the ability to identify defaulters (higher recall) but also maintain precision without a significant decline by effectively controlling misclassifications (i.e., false positives). Moreover, F1 scores for ChatGPT-refined models are generally higher than those of human-written models. In sum, these findings, based on more realistic rejection thresholds, are consistent with our main experimental results using AUC, KS, and H-measure, which further reinforce the validity and practicality of our proposed framework.

**Table F.1** Model performance of human-written and ChatGPT-refined texts (profit-based threshold, 70 rejections, and 100 rejections)

| Model | Category | Profit-based threshold | | | 70 rejections | | | 100 rejections | | |
|---|---|---|---|---|---|---|---|---|---|---|
| | | Structured | Text | Combined | Structured | Text | Combined | Structured | Text | Combined |
| | | Recall | | | Recall | | | Recall | | |
| LDA+MLP | Human | 0.115 (0.111, 0.118) | 0.170 (0.168, 0.173) | **0.274 (0.271, 0.277)** | 0.273 (0.269, 0.277) | 0.220 (0.217, 0.223) | 0.280 (0.277, 0.283) | 0.311 (0.308, 0.315) | 0.229 (0.226, 0.231) | 0.305 (0.302, 0.308) |
| | ChatGPT | 0.115 (0.111, 0.118) | 0.035 (0.033, 0.037) | 0.260 (0.256, 0.263) | 0.273 (0.269, 0.277) | 0.108 (0.105, 0.110) | 0.294 (0.291, 0.297) | 0.311 (0.308, 0.315) | 0.177 (0.173, 0.180) | 0.348 (0.345, 0.351) |
| fastText+MLP | Human | 0.115 (0.111, 0.118) | 0.166 (0.164, 0.168) | 0.259 (0.256, 0.262) | 0.273 (0.269, 0.277) | 0.209 (0.206, 0.211) | 0.330 (0.327, 0.333) | 0.311 (0.308, 0.315) | 0.221 (0.218, 0.224) | 0.336 (0.333, 0.339) |
| | ChatGPT | 0.115 (0.111, 0.118) | 0.135 (0.132, 0.139) | 0.271 (0.268, 0.275) | 0.273 (0.269, 0.277) | 0.305 (0.301, 0.309) | 0.335 (0.332, 0.338) | 0.311 (0.308, 0.315) | 0.336 (0.332, 0.341) | 0.367 (0.364, 0.370) |
| Ada-002+MLP | Human | 0.115 (0.111, 0.118) | 0.229 (0.226, 0.232) | 0.221 (0.219, 0.224) | 0.273 (0.269, 0.277) | 0.386 (0.383, 0.389) | 0.306 (0.303, 0.309) | 0.311 (0.308, 0.315) | **0.443 (0.440, 0.447)** | 0.386 (0.382, 0.389) |
| | ChatGPT | 0.115 (0.111, 0.118) | **0.327 (0.323, 0.331)** | 0.264 (0.261, 0.267) | 0.273 (0.269, 0.277) | **0.390 (0.387, 0.394)** | 0.286 (0.283, 0.289) | 0.311 (0.308, 0.315) | 0.390 (0.387, 0.394) | 0.372 (0.369, 0.375) |
| BERT+MLP | Human | 0.115 (0.111, 0.118) | 0.194 (0.191, 0.197) | 0.223 (0.220, 0.226) | 0.273 (0.269, 0.277) | 0.265 (0.262, 0.269) | **0.333 (0.329, 0.336)** | 0.311 (0.308, 0.315) | 0.301 (0.298, 0.305) | 0.382 (0.378, 0.386) |
| | ChatGPT | 0.115 (0.111, 0.118) | 0.135 (0.132, 0.138) | 0.185 (0.181, 0.190) | 0.273 (0.269, 0.277) | 0.244 (0.241, 0.247) | 0.308 (0.305, 0.312) | 0.311 (0.308, 0.315) | 0.305 (0.302, 0.308) | **0.402 (0.398, 0.407)** |
| | | Precision | | | Precision | | | Precision | | |
| LDA+MLP | Human | 0.208 (0.201, 0.215) | 0.413 (0.406, 0.420) | 0.394 (0.390, 0.398) | 0.069 (0.068, 0.070) | 0.055 (0.055, 0.056) | 0.070 (0.070, 0.071) | 0.055 (0.055, 0.056) | 0.041 (0.040, 0.041) | 0.054 (0.054, 0.055) |
| | ChatGPT | 0.208 (0.201, 0.215) | 0.060 (0.058, 0.063) | 0.254 (0.252, 0.257) | 0.069 (0.068, 0.070) | 0.027 (0.026, 0.028) | 0.074 (0.073, 0.075) | 0.055 (0.055, 0.056) | 0.031 (0.031, 0.032) | 0.062 (0.061, 0.062) |
| fastText+MLP | Human | 0.208 (0.201, 0.215) | **0.459 (0.452, 0.467)** | 0.358 (0.353, 0.363) | 0.069 (0.068, 0.070) | 0.053 (0.052, 0.053) | 0.083 (0.082, 0.084) | 0.055 (0.055, 0.056) | 0.039 (0.039, 0.040) | 0.060 (0.059, 0.060) |
| | ChatGPT | 0.208 (0.201, 0.215) | 0.186 (0.182, 0.191) | 0.308 (0.304, 0.311) | 0.069 (0.068, 0.070) | 0.077 (0.076, 0.078) | **0.084 (0.084, 0.085)** | 0.055 (0.055, 0.056) | 0.060 (0.059, 0.060) | 0.065 (0.065, 0.066) |
| Ada-002+MLP | Human | 0.208 (0.201, 0.215) | 0.430 (0.425, 0.436) | **0.433 (0.427, 0.438)** | 0.069 (0.068, 0.070) | 0.097 (0.096, 0.098) | 0.077 (0.076, 0.078) | 0.055 (0.055, 0.056) | **0.079 (0.078, 0.079)** | 0.068 (0.068, 0.069) |
| | ChatGPT | 0.208 (0.201, 0.215) | 0.250 (0.247, 0.254) | 0.299 (0.295, 0.302) | 0.069 (0.068, 0.070) | **0.098 (0.097, 0.099)** | 0.072 (0.071, 0.073) | 0.055 (0.055, 0.056) | 0.069 (0.069, 0.070) | 0.066 (0.065, 0.067) |
| BERT+MLP | Human | 0.208 (0.201, 0.215) | 0.389 (0.382, 0.395) | 0.356 (0.351, 0.362) | 0.069 (0.068, 0.070) | 0.067 (0.066, 0.068) | 0.084 (0.083, 0.085) | 0.055 (0.055, 0.056) | 0.053 (0.053, 0.054) | 0.068 (0.067, 0.068) |
| | ChatGPT | 0.208 (0.201, 0.215) | 0.188 (0.184, 0.193) | 0.202 (0.198, 0.207) | 0.069 (0.068, 0.070) | 0.061 (0.061, 0.062) | 0.078 (0.077, 0.079) | 0.055 (0.055, 0.056) | 0.054 (0.054, 0.055) | **0.071 (0.071, 0.072)** |
| | | F1 | | | F1 | | | F1 | | |
| LDA+MLP | Human | 0.121 (0.118, 0.124) | 0.229 (0.226, 0.232) | **0.315 (0.311, 0.318)** | 0.110 (0.108, 0.111) | 0.089 (0.087, 0.090) | 0.113 (0.111, 0.114) | 0.094 (0.093, 0.095) | 0.069 (0.068, 0.070) | 0.092 (0.091, 0.093) |
| | ChatGPT | 0.121 (0.118, 0.124) | 0.042 (0.040, 0.044) | 0.250 (0.247, 0.253) | 0.110 (0.108, 0.111) | 0.043 (0.042, 0.044) | 0.118 (0.117, 0.119) | 0.094 (0.093, 0.095) | 0.053 (0.052, 0.054) | 0.105 (0.104, 0.106) |
| fastText+MLP | Human | 0.121 (0.118, 0.124) | 0.229 (0.225, 0.232) | 0.286 (0.283, 0.289) | 0.110 (0.108, 0.111) | 0.084 (0.083, 0.085) | 0.133 (0.131, 0.134) | 0.094 (0.093, 0.095) | 0.067 (0.066, 0.067) | 0.101 (0.100, 0.102) |
| | ChatGPT | 0.121 (0.118, 0.124) | 0.146 (0.142, 0.149) | 0.281 (0.278, 0.284) | 0.110 (0.108, 0.111) | 0.123 (0.121, 0.124) | **0.135 (0.134, 0.136)** | 0.094 (0.093, 0.095) | 0.101 (0.100, 0.103) | 0.111 (0.110, 0.111) |
| Ada-002+MLP | Human | 0.121 (0.118, 0.124) | **0.283 (0.280, 0.286)** | 0.283 (0.279, 0.286) | 0.110 (0.108, 0.111) | 0.155 (0.154, 0.157) | 0.123 (0.122, 0.124) | 0.094 (0.093, 0.095) | **0.133 (0.132, 0.135)** | 0.116 (0.115, 0.117) |
| | ChatGPT | 0.121 (0.118, | 0.266 (0.264, | 0.273 (0.270, | 0.110 (0.108, | **0.157 (0.156,** | 0.115 (0.114, | 0.094 (0.093, | 0.118 (0.117, | 0.112 (0.111, |

|  |  |  |  |  |  |  |  |  |  |
|---|---|---|---|---|---|---|---|---|---|
|  |  | 0.124) | 0.269) | 0.275) | 0.111) | **0.158)** | 0.116) | 0.095) | 0.119) | 0.113) |
| BERT+MLP | Human | 0.121 (0.118, 0.124) | 0.238 (0.234, 0.241) | 0.255 (0.252, 0.258) | 0.110 (0.108, 0.111) | 0.107 (0.105, 0.108) | 0.134 (0.133, 0.135) | 0.094 (0.093, 0.095) | 0.091 (0.090, 0.092) | 0.115 (0.114, 0.116) |
|  | ChatGPT | 0.121 (0.118, 0.124) | 0.142 (0.139, 0.145) | 0.171 (0.168, 0.175) | 0.110 (0.108, 0.111) | 0.098 (0.097, 0.099) | 0.124 (0.123, 0.126) | 0.094 (0.093, 0.095) | 0.092 (0.091, 0.093) | **0.121 (0.120, 0.123)** |

**Notes:** This table presents the discrimination performance of the recall, precision, and F1 scores (mean and its 95% confidence interval) for predicting the default risk of borrowers based on four NLP approaches: LDA+MLP, fastText+MLP, Ada-002+MLP, and BERT+MLP.

**Table F.2** Model performance of human-written and ChatGPT-refined texts (120 rejections, 150 rejections, and 165 rejections)

| Model | Category | 120 rejections | | | 150 rejections | | | 165 rejections | | |
|---|---|---|---|---|---|---|---|---|---|---|
|  |  | Structured | Text | Combined | Structured | Text | Combined | Structured | Text | Combined |
|  |  | Recall | | | Recall | | | Recall | | |
| LDA+MLP | Human | 0.340 (0.337, 0.344) | 0.233 (0.230, 0.236) | 0.351 (0.348, 0.354) | 0.366 (0.362, 0.370) | 0.257 (0.255, 0.260) | 0.423 (0.419, 0.426) | 0.381 (0.377, 0.385) | 0.286 (0.283, 0.289) | 0.436 (0.432, 0.439) |
|  | ChatGPT | 0.340 (0.337, 0.344) | 0.258 (0.255, 0.262) | 0.414 (0.411, 0.418) | 0.366 (0.362, 0.370) | 0.385 (0.381, 0.389) | 0.444 (0.440, 0.447) | 0.381 (0.377, 0.385) | 0.419 (0.415, 0.423) | 0.446 (0.443, 0.450) |
| fastText+MLP | Human | 0.340 (0.337, 0.344) | 0.221 (0.219, 0.224) | 0.336 (0.333, 0.339) | 0.366 (0.362, 0.370) | 0.246 (0.243, 0.249) | 0.337 (0.334, 0.340) | 0.381 (0.377, 0.385) | 0.280 (0.277, 0.283) | 0.347 (0.344, 0.350) |
|  | ChatGPT | 0.340 (0.337, 0.344) | 0.363 (0.359, 0.367) | 0.433 (0.429, 0.436) | 0.366 (0.362, 0.370) | 0.411 (0.408, 0.415) | 0.447 (0.443, 0.450) | 0.381 (0.377, 0.385) | 0.441 (0.437, 0.445) | 0.447 (0.443, 0.450) |
| Ada-002+MLP | Human | 0.340 (0.337, 0.344) | **0.446 (0.442, 0.449)** | 0.406 (0.402, 0.409) | 0.366 (0.362, 0.370) | **0.446 (0.442, 0.449)** | 0.489 (0.486, 0.492) | 0.381 (0.377, 0.385) | 0.446 (0.442, 0.449) | 0.501 (0.498, 0.504) |
|  | ChatGPT | 0.340 (0.337, 0.344) | 0.391 (0.387, 0.394) | 0.436 (0.432, 0.439) | 0.366 (0.362, 0.370) | 0.425 (0.421, 0.428) | 0.460 (0.457, 0.460) | 0.381 (0.377, 0.385) | **0.470 (0.466, 0.473)** | 0.487 (0.484, 0.491) |
| BERT+MLP | Human | 0.340 (0.337, 0.344) | 0.339 (0.335, 0.342) | 0.400 (0.396, 0.404) | 0.366 (0.362, 0.370) | 0.372 (0.368, 0.375) | 0.448 (0.444, 0.451) | 0.381 (0.377, 0.385) | 0.387 (0.383, 0.390) | 0.483 (0.479, 0.486) |
|  | ChatGPT | 0.340 (0.337, 0.344) | 0.332 (0.329, 0.336) | **0.453 (0.448, 0.459)** | 0.366 (0.362, 0.370) | 0.371 (0.368, 0.374) | **0.511 (0.506, 0.516)** | 0.381 (0.377, 0.385) | 0.389 (0.385, 0.392) | **0.539 (0.535, 0.544)** |
|  |  | Precision | | | Precision | | | Precision | | |
| LDA+MLP | Human | 0.050 (0.050, 0.051) | 0.035 (0.034, 0.035) | 0.052 (0.052, 0.052) | 0.044 (0.043, 0.044) | 0.031 (0.030, 0.031) | 0.050 (0.050, 0.051) | 0.041 (0.041, 0.042) | 0.031 (0.031, 0.031) | 0.047 (0.047, 0.047) |
|  | ChatGPT | 0.050 (0.050, 0.051) | 0.038 (0.038, 0.039) | 0.061 (0.061, 0.062) | 0.044 (0.043, 0.044) | 0.046 (0.045, 0.046) | 0.053 (0.052, 0.053) | 0.041 (0.041, 0.042) | 0.045 (0.045, 0.046) | 0.048 (0.048, 0.049) |
| fastText+MLP | Human | 0.050 (0.050, 0.051) | 0.033 (0.032, 0.033) | 0.050 (0.049, 0.050) | 0.044 (0.043, 0.044) | 0.029 (0.029, 0.030) | 0.040 (0.040, 0.040) | 0.041 (0.041, 0.042) | 0.030 (0.030, 0.031) | 0.037 (0.037, 0.038) |

|  |  |  |  |  |  |  |  |  |  |
|---|---|---|---|---|---|---|---|---|---|
| LDA+MLP | Human | 0.050 (0.050, 0.051) | 0.050 (0.050, 0.051) | 0.063 (0.062, 0.064) | 0.044 (0.043, 0.044) | 0.044 (0.044, 0.045) | 0.060 (0.060, 0.061) | 0.041 (0.041, 0.042) | 0.042 (0.042, 0.043) | 0.055 (0.054, 0.056) |
|  | ChatGPT | 0.050 (0.050, 0.051) | 0.054 (0.053, 0.054) | 0.069 (0.068, 0.070) | 0.044 (0.043, 0.044) | 0.049 (0.048, 0.049) | 0.057 (0.057, 0.058) | 0.041 (0.041, 0.042) | 0.048 (0.047, 0.048) | 0.052 (0.052, 0.053) |
| fastText+ MLP | Human | 0.050 (0.050, 0.051) | 0.048 (0.047, 0.048) | 0.060 (0.059, 0.061) | 0.044 (0.043, 0.044) | 0.044 (0.044, 0.045) | 0.050 (0.050, 0.051) | 0.041 (0.041, 0.042) | 0.042 (0.042, 0.043) | 0.046 (0.045, 0.046) |
|  | ChatGPT | 0.050 (0.050, 0.051) | 0.054 (0.053, 0.054) | 0.064 (0.064, 0.065) | 0.044 (0.043, 0.044) | 0.049 (0.048, 0.049) | 0.053 (0.053, 0.053) | 0.041 (0.041, 0.042) | 0.048 (0.047, 0.048) | 0.048 (0.048, 0.049) |
| Ada-002+ MLP | Human | 0.050 (0.050, 0.051) | **0.066 (0.065, 0.067)** | 0.060 (0.060, 0.061) | 0.044 (0.043, 0.044) | **0.053 (0.053, 0.053)** | 0.058 (0.058, 0.058) | 0.041 (0.041, 0.042) | 0.048 (0.048, 0.049) | 0.054 (0.054, 0.055) |
|  | ChatGPT | 0.050 (0.050, 0.051) | 0.058 (0.057, 0.058) | 0.065 (0.064, 0.065) | 0.044 (0.043, 0.044) | 0.050 (0.050, 0.051) | 0.055 (0.054, 0.055) | 0.041 (0.041, 0.042) | **0.051 (0.050, 0.051)** | 0.053 (0.052, 0.053) |
| BERT+MLP | Human | 0.050 (0.050, 0.051) | 0.050 (0.050, 0.051) | 0.059 (0.059, 0.060) | 0.044 (0.043, 0.044) | 0.044 (0.044, 0.045) | 0.053 (0.053, 0.054) | 0.041 (0.041, 0.042) | 0.042 (0.041, 0.042) | 0.052 (0.052, 0.053) |
|  | ChatGPT | 0.050 (0.050, 0.051) | 0.049 (0.049, 0.050) | **0.067 (0.066, 0.068)** | 0.044 (0.043, 0.044) | 0.044 (0.044, 0.044) | **0.061 (0.060, 0.061)** | 0.041 (0.041, 0.042) | 0.042 (0.042, 0.042) | **0.058 (0.058, 0.059)** |
|  |  | F1 | | | F1 | | | F1 | | |
| LDA+MLP | Human | 0.088 (0.087, 0.089) | 0.060 (0.059, 0.061) | 0.091 (0.090, 0.091) | 0.078 (0.077, 0.079) | 0.055 (0.054, 0.055) | 0.090 (0.089, 0.090) | 0.074 (0.074, 0.075) | 0.056 (0.055, 0.056) | 0.085 (0.084, 0.086) |
|  | ChatGPT | 0.088 (0.087, 0.089) | 0.067 (0.066, 0.068) | 0.107 (0.106, 0.108) | 0.078 (0.077, 0.079) | 0.082 (0.081, 0.083) | 0.094 (0.093, 0.095) | 0.074 (0.074, 0.075) | 0.082 (0.081, 0.083) | 0.087 (0.086, 0.088) |
| fastText+ MLP | Human | 0.088 (0.087, 0.089) | 0.057 (0.056, 0.058) | 0.087 (0.086, 0.088) | 0.078 (0.077, 0.079) | 0.052 (0.052, 0.053) | 0.072 (0.071, 0.072) | 0.074 (0.074, 0.075) | 0.055 (0.054, 0.055) | 0.068 (0.067, 0.068) |
|  | ChatGPT | 0.088 (0.087, 0.089) | 0.094 (0.093, 0.095) | 0.112 (0.111, 0.113) | 0.078 (0.077, 0.079) | 0.087 (0.087, 0.088) | 0.095 (0.094, 0.096) | 0.074 (0.074, 0.075) | 0.086 (0.085, 0.087) | 0.087 (0.086, 0.088) |
| Ada-002+ MLP | Human | 0.088 (0.087, 0.089) | **0.115 (0.114, 0.116)** | 0.105 (0.104, 0.106) | 0.078 (0.077, 0.079) | **0.095 (0.094, 0.095)** | 0.104 (0.103, 0.105) | 0.074 (0.074, 0.075) | 0.087 (0.086, 0.088) | 0.098 (0.097, 0.098) |
|  | ChatGPT | 0.088 (0.087, 0.089) | 0.101 (0.100, 0.102) | 0.112 (0.112, 0.113) | 0.078 (0.077, 0.079) | 0.090 (0.089, 0.091) | 0.098 (0.097, 0.098) | 0.074 (0.074, 0.075) | **0.092 (0.091, 0.092)** | 0.095 (0.094, 0.096) |
| BERT+MLP | Human | 0.088 (0.087, 0.089) | 0.087 (0.087, 0.088) | 0.103 (0.102, 0.104) | 0.078 (0.077, 0.079) | 0.079 (0.078, 0.080) | 0.095 (0.094, 0.096) | 0.074 (0.074, 0.075) | 0.075 (0.075, 0.076) | 0.094 (0.094, 0.095) |
|  | ChatGPT | 0.088 (0.087, 0.089) | 0.086 (0.085, 0.087) | **0.117 (0.116, 0.118)** | 0.078 (0.077, 0.079) | 0.079 (0.078, 0.080) | **0.108 (0.107, 0.110)** | 0.074 (0.074, 0.075) | 0.076 (0.075, 0.077) | **0.105 (0.104, 0.106)** |

**Notes:** This table presents the discrimination performance of the recall, precision, and F1 scores (mean and its 95% confidence interval) for predicting the default risk of borrowers based on four NLP approaches: LDA+MLP, fastText+MLP, Ada-002+MLP, and BERT+MLP.

# Appendix G. Robustness check across alternative classifiers

**Table G.1** Search space of hyperparameters for alternative classifiers

| Classifier | Search space |
|---|---|
| LR | $C \in \{0.01, 0.1, 1\}$ |
| DT | $max\_depth \in \{2, 6, 8, 10\}$ |
|  | $min\_samples\_split \in \{2, 6, 8, 10\}$ |
|  | $min\_samples\_leaf \in \{2, 6, 8, 10\}$ |
| RF | $max\_depth \in \{2, 6, 8, 10\}$ |
|  | $min\_samples\_split \in \{2, 6, 8, 10\}$ |
|  | $min\_samples\_leaf \in \{2, 6, 8, 10\}$ |
|  | $n\_estimators \in \{30, 70, 90\}$ |
| SVM | $C \in \{0.1, 1, 2, 3, 4\}$ |
|  | $class\_weight \in \{None, "balanced"\}$ |
|  | $kernel \in \{"rbf", "poly"\}$ |
|  | $gamma \in \{"scale", "auto"\}$ |

**Notes:** This table shows the considered hyperparameter spaces for the alternative classifiers used for the robustness check of the results in Section 5.1 in our study. The final hyperparameter combination is chosen based on the best validation set performance.

**Table G.2** Model performance of human-written and ChatGPT-refined texts on alternative classifiers

| | | Human | | | ChatGPT | |
|---|---|---|---|---|---|---|
| NLP method | Classifier | Structure | Text | Combined | Text | Combined |
| | | | | AUC | | |
| LDA | LR | 0.665 (0.663, 0.667) | 0.592 (0.590, 0.594) | 0.668 (0.666, 0.670) | 0.716 (0.714, 0.717) | 0.711 (0.710, 0.713) |
| | DT | 0.589 (0.588, 0.591) | 0.603 (0.601, 0.604) | 0.617 (0.615, 0.618) | 0.659 (0.657, 0.661) | 0.629 (0.627, 0.630) |
| | RF | 0.694 (0.692, 0.696) | 0.665 (0.663, 0.667) | 0.692 (0.689, 0.694) | 0.684 (0.683, 0.686) | 0.701 (0.699, 0.703) |
| | SVM | 0.671 (0.669, 0.673) | 0.670 (0.668, 0.671) | 0.653 (0.651, 0.655) | 0.726 (0.725, 0.728) | 0.688 (0.686, 0.689) |
| fastText | LR | 0.665 (0.663, 0.667) | 0.549 (0.547, 0.551) | 0.666 (0.664, 0.668) | 0.679 (0.677, 0.681) | 0.672 (0.670, 0.674) |
| | DT | 0.589 (0.588, 0.591) | 0.598 (0.597, 0.600) | 0.599 (0.598, 0.600) | 0.598 (0.596, 0.600) | 0.598 (0.596, 0.600) |
| | RF | 0.694 (0.692, 0.696) | 0.670 (0.668, 0.672) | **0.708 (0.706, 0.710)** | 0.615 (0.613, 0.617) | 0.582 (0.579, 0.584) |
| | SVM | 0.671 (0.669, 0.673) | **0.693 (0.691, 0.695)** | 0.659 (0.658, 0.661) | 0.648 (0.645, 0.650) | 0.628 (0.626, 0.630) |
| Ada-002 | LR | 0.665 (0.663, 0.667) | 0.648 (0.645, 0.650) | 0.676 (0.674, 0.678) | **0.732 (0.730, 0.734)** | 0.671 (0.669, 0.672) |
| | DT | 0.589 (0.588, 0.591) | 0.629 (0.627, 0.631) | 0.625 (0.623, 0.627) | 0.656 (0.654, 0.657) | 0.656 (0.655, 0.658) |
|  | RF | 0.694 (0.692, 0.696) | 0.608 (0.605, 0.610) | 0.622 (0.619, 0.624) | 0.654 (0.652, 0.656) | 0.642 (0.639, 0.644) |
|  | SVM | 0.671 (0.669, 0.673) | 0.594 (0.592, 0.596) | 0.640 (0.638, 0.642) | 0.658 (0.655, 0.660) | 0.632 (0.630, 0.634) |
| BERT | LR | 0.665 (0.663, 0.667) | 0.598 (0.596, 0.600) | 0.678 (0.677, 0.680) | 0.694 (0.692, 0.696) | **0.734 (0.732, 0.735)** |
|  | DT | 0.589 (0.588, 0.591) | 0.581 (0.579, 0.582) | 0.595 (0.593, 0.596) | 0.578 (0.576, 0.579) | 0.540 (0.538, 0.541) |
|  | RF | 0.694 (0.692, 0.696) | 0.624 (0.622, 0.627) | 0.658 (0.656, 0.660) | 0.686 (0.684, 0.689) | 0.715 (0.712, 0.717) |
|  | SVM | 0.671 (0.669, 0.673) | 0.574 (0.572, 0.576) | 0.678 (0.677, 0.680) | 0.666 (0.664, 0.668) | 0.700 (0.699, 0.702) |
|  |  | KS |  |  |  |  |
| LDA | LR | 0.343 (0.341, 0.345) | 0.264 (0.262, 0.267) | 0.350 (0.348, 0.352) | 0.438 (0.436, 0.441) | 0.419 (0.416, 0.421) |
|  | DT | 0.231 (0.228, 0.234) | 0.244 (0.241, 0.247) | 0.249 (0.246, 0.252) | 0.295 (0.291, 0.298) | 0.268 (0.265, 0.272) |
|  | RF | 0.444 (0.441, 0.447) | 0.354 (0.352, 0.357) | **0.418 (0.415, 0.421)** | 0.432 (0.430, 0.435) | 0.400 (0.397, 0.402) |
|  | SVM | 0.357 (0.354, 0.359) | 0.343 (0.340, 0.345) | 0.340 (0.338, 0.343) | 0.438 (0.436, 0.440) | 0.372 (0.370, 0.375) |
| fastText | LR | 0.343 (0.341, 0.345) | 0.226 (0.223, 0.228) | 0.343 (0.341, 0.346) | 0.413 (0.410, 0.416) | 0.356 (0.353, 0.358) |
|  | DT | 0.231 (0.228, 0.234) | 0.205 (0.202, 0.207) | 0.205 (0.202, 0.207) | 0.217 (0.214, 0.219) | 0.217 (0.214, 0.219) |
|  | RF | 0.444 (0.441, 0.447) | 0.374 (0.371, 0.377) | 0.406 (0.403, 0.409) | 0.295 (0.292, 0.297) | 0.285 (0.282, 0.288) |
|  | SVM | 0.357 (0.354, 0.359) | **0.434 (0.431, 0.437)** | 0.334 (0.332, 0.337) | 0.399 (0.396, 0.402) | 0.287 (0.285, 0.290) |
| Ada-002 | LR | 0.343 (0.341, 0.345) | 0.380 (0.377, 0.383) | 0.357 (0.355, 0.359) | **0.481 (0.478, 0.484)** | 0.350 (0.347, 0.352) |
|  | DT | 0.231 (0.228, 0.234) | 0.261 (0.258, 0.264) | 0.253 (0.250, 0.256) | 0.301 (0.298, 0.303) | 0.301 (0.298, 0.304) |
|  | RF | 0.444 (0.441, 0.447) | 0.329 (0.327, 0.332) | 0.329 (0.326, 0.331) | 0.348 (0.345, 0.351) | 0.336 (0.333, 0.339) |
|  | SVM | 0.357 (0.354, 0.359) | 0.283 (0.280, 0.286) | 0.317 (0.314, 0.319) | 0.429 (0.426, 0.433) | 0.297 (0.295, 0.300) |
| BERT | LR | 0.343 (0.341, 0.345) | 0.328 (0.325, 0.331) | 0.355 (0.353, 0.358) | 0.443 (0.440, 0.445) | **0.488 (0.486, 0.490)** |
|  | DT | 0.231 (0.228, 0.234) | 0.182 (0.179, 0.184) | 0.198 (0.195, 0.201) | 0.157 (0.154, 0.159) | 0.091 (0.088, 0.093) |
|  | RF | 0.444 (0.441, 0.447) | 0.320 (0.317, 0.323) | 0.358 (0.355, 0.360) | 0.408 (0.405, 0.411) | 0.440 (0.436, 0.443) |
|  | SVM | 0.357 (0.354, 0.359) | 0.263 (0.260, 0.266) | 0.371 (0.369, 0.374) | 0.353 (0.350, 0.355) | 0.389 (0.387, 0.391) |
|  |  | H-measure |  |  |  |  |
| LDA | LR | 0.221 (0.219, 0.223) | 0.177 (0.175, 0.179) | 0.263 (0.261, 0.266) | 0.231 (0.229, 0.234) | 0.267 (0.265, 0.269) |
|  | DT | 0.176 (0.174, 0.179) | 0.190 (0.187, 0.193) | 0.118 (0.116, 0.121) | 0.121 (0.119, 0.123) | 0.090 (0.088, 0.091) |
|  | RF | 0.299 (0.296, 0.302) | 0.252 (0.250, 0.255) | **0.302 (0.299, 0.305)** | 0.201 (0.199, 0.203) | 0.252 (0.250, 0.255) |
|  | SVM | 0.261 (0.258, 0.264) | 0.272 (0.270, 0.275) | 0.266 (0.264, 0.269) | 0.243 (0.241, 0.246) | 0.239 (0.237, 0.242) |
| fastText | LR | 0.221 (0.219, 0.223) | 0.169 (0.167, 0.172) | 0.274 (0.271, 0.276) | 0.281 (0.279, 0.284) | 0.230 (0.228, 0.232) |
|  | DT | 0.176 (0.174, 0.179) | 0.155 (0.152, 0.157) | 0.155 (0.152, 0.157) | 0.053 (0.052, 0.054) | 0.053 (0.052, 0.054) |
|  | RF | 0.299 (0.296, 0.302) | 0.263 (0.260, 0.265) | 0.284 (0.282, 0.287) | 0.196 (0.194, 0.198) | 0.200 (0.198, 0.202) |
|  | SVM | 0.261 (0.258, 0.264) | **0.312 (0.309, 0.315)** | 0.218 (0.216, 0.220) | 0.318 (0.315, 0.321) | 0.174 (0.172, 0.176) |
| Ada-002 | LR  | 0.221 (0.219, 0.223) | 0.295 (0.292, 0.298) | 0.272 (0.269, 0.274) | **0.365 (0.362, 0.368)** | 0.225 (0.223, 0.227)     |
|      | DT  | 0.176 (0.174, 0.179) | 0.147 (0.144, 0.150) | 0.142 (0.140, 0.145) | 0.087 (0.086, 0.089)     | 0.088 (0.086, 0.089)     |
|      | RF  | 0.299 (0.296, 0.302) | 0.267 (0.264, 0.269) | 0.269 (0.266, 0.271) | 0.243 (0.240, 0.245)     | 0.221 (0.219, 0.224)     |
|      | SVM | 0.261 (0.258, 0.264) | 0.227 (0.225, 0.230) | 0.196 (0.194, 0.198) | 0.341 (0.338, 0.344)     | 0.180 (0.177, 0.182)     |
| BERT | LR  | 0.221 (0.219, 0.223) | 0.244 (0.242, 0.247) | 0.272 (0.270, 0.275) | 0.253 (0.251, 0.256)     | 0.321 (0.318, 0.323)     |
|      | DT  | 0.176 (0.174, 0.179) | 0.134 (0.132, 0.136) | 0.141 (0.138, 0.143) | 0.070 (0.068, 0.072)     | 0.025 (0.024, 0.026)     |
|      | RF  | 0.299 (0.296, 0.302) | 0.240 (0.237, 0.242) | 0.264 (0.261, 0.266) | 0.305 (0.302, 0.308)     | **0.332 (0.329, 0.335)** |
|      | SVM | 0.261 (0.258, 0.264) | 0.211 (0.208, 0.213) | 0.278 (0.275, 0.280) | 0.260 (0.258, 0.263)     | 0.265 (0.263, 0.268)     |
| PRAUC ||||||
| LDA  | LR  | 0.103 (0.102, 0.105) | 0.079 (0.078, 0.081) | 0.138 (0.136, 0.140) | 0.051 (0.050, 0.051)     | 0.103 (0.102, 0.104)     |
|      | DT  | 0.107 (0.105, 0.109) | 0.206 (0.204, 0.209) | 0.049 (0.049, 0.050) | 0.181 (0.179, 0.183)     | 0.148 (0.146, 0.149)     |
|      | RF  | 0.143 (0.141, 0.145) | 0.242 (0.239, 0.245) | 0.268 (0.266, 0.271) | 0.046 (0.045, 0.046)     | 0.117 (0.115, 0.119)     |
|      | SVM | 0.124 (0.122, 0.126) | 0.271 (0.268, 0.274) | 0.106 (0.105, 0.108) | 0.054 (0.053, 0.054)     | 0.090 (0.088, 0.091)     |
| fastText | LR  | 0.103 (0.102, 0.105) | 0.193 (0.191, 0.196) | 0.258 (0.255, 0.261) | 0.100 (0.099, 0.101)     | 0.096 (0.094, 0.097)     |
|      | DT  | 0.107 (0.105, 0.109) | 0.211 (0.209, 0.213) | 0.214 (0.211, 0.216) | 0.246 (0.244, 0.248)     | **0.246 (0.244, 0.248)** |
|      | RF  | 0.143 (0.141, 0.145) | 0.250 (0.247, 0.253) | 0.263 (0.260, 0.266) | 0.173 (0.171, 0.176)     | 0.176 (0.173, 0.178)     |
|      | SVM | 0.124 (0.122, 0.126) | 0.272 (0.269, 0.275) | 0.079 (0.078, 0.080) | 0.209 (0.206, 0.212)     | 0.059 (0.058, 0.060)     |
| Ada-002 | LR  | 0.103 (0.102, 0.105) | 0.251 (0.248, 0.254) | 0.153 (0.151, 0.155) | 0.184 (0.181, 0.186)     | 0.096 (0.094, 0.097)     |
|      | DT  | 0.107 (0.105, 0.109) | 0.056 (0.055, 0.056) | 0.056 (0.055, 0.056) | 0.088 (0.084, 0.092)     | 0.112 (0.107, 0.116)     |
|      | RF  | 0.143 (0.141, 0.145) | **0.281 (0.278, 0.284)** | **0.282 (0.279, 0.285)** | 0.150 (0.148, 0.153) | 0.157 (0.154, 0.159)     |
|      | SVM | 0.124 (0.122, 0.126) | 0.205 (0.202, 0.207) | 0.058 (0.057, 0.058) | **0.282 (0.279, 0.285)** | 0.053 (0.052, 0.053)     |
| BERT | LR  | 0.103 (0.102, 0.105) | 0.207 (0.204, 0.209) | 0.245 (0.242, 0.248) | 0.179 (0.177, 0.182)     | 0.129 (0.127, 0.131)     |
|      | DT  | 0.107 (0.105, 0.109) | 0.151 (0.149, 0.153) | 0.129 (0.127, 0.131) | 0.068 (0.067, 0.070)     | 0.059 (0.057, 0.060)     |
|      | RF  | 0.143 (0.141, 0.145) | 0.230 (0.227, 0.233) | 0.242 (0.239, 0.245) | 0.205 (0.202, 0.207)     | 0.228 (0.225, 0.231)     |
|      | SVM | 0.124 (0.122, 0.126) | 0.228 (0.225, 0.231) | 0.087 (0.086, 0.088) | 0.193 (0.190, 0.195)     | 0.114 (0.113, 0.116)     |

**Notes**: This table presents the performance results of the AUC, KS scores, H-measures, and PRAUC (mean and its 95% confidence interval) for predicting the default risk of borrowers based on four NLP approaches and four classifiers including LR, DT, RF and SVM. For the BERT-combined models, the classifiers are trained using the integration of a 768-dimensional pooled representation produced by BERT with the structured data.

In Table G.2, we observe that the combined models generally display superior performance to the text-only models. Furthermore, most models relying on ChatGPT-refined texts surpass those using human-written texts in terms of AUC and KS, but the opposite trend is observed for H-measure and PRAUC. When comparing classifiers, we find that LR, RF, and SVM outperform DT in most cases. Notably, the fastText+RF and BERT+LR combined models achieve the highest average AUC scores of 0.708 and 0.734 for the human-written and ChatGPT-refined texts, respectively. When it comes to the effectiveness of textual feature extraction, the results of using the LDA model are comparable to those obtained using the fastText, Ada-002, and BERT models. This may be due to the fact that fastText, Ada-002, and BERT derive 300-, 1536-, and 768-dimensional features from texts, respectively, thus causing the classifiers to suffer from the curse of dimensionality. In summary, these results bolster our confidence in the conclusions drawn from our primary analysis. Specifically, integrating unstructured text data can significantly enhance credit default predictions, and ChatGPT-refined texts enhance the predictive power of credit scoring models in terms of AUC and KS, but their impact on H-measure and PRAUC remains mixed.